\documentclass[10pt,twoside]{article}
\usepackage{a4wide,amssymb,epsfig,tildes}
\usepackage[tight]{subfigure}
\usepackage{amsmath,fancyhdr}

\date{}

\title{\textbf{Geometric modelling of kink banding}\\ \textbf{in
    laminated structures}}

\author{\normalsize\textsc{By M. Ahmer Wadee}$^{1,}$\footnote{Author for
    correspondence: a.wadee@imperial.ac.uk}, \textsc{Christina V\"ollmecke}$^2$,
  \textsc{Joseph F. Haley}$^1$\\ \normalsize \textsc{and Stylianos
    Yiatros}$^3$\\~\\
  \normalsize $^1$\emph{Department of Civil \& Environmental
    Engineering, Imperial College London, London, UK}\\
  \normalsize $^2$\emph{LKM, Institut f\"ur Mechanik, Technische Universit\"at
    Berlin, Germany}\\
  \normalsize $^3$\emph{Department of Civil Engineering, Brunel University,
    Uxbridge, UK}}

\DeclareMathOperator{\sgn}{sgn}

\begin{document}

\thispagestyle{fancy}
\renewcommand{\headrulewidth}{0pt}
\renewcommand\footnoterule{}

\maketitle

\begin{abstract}
  \noindent
  An analytical model founded on geometric and potential energy
  principles for kink band deformation in laminated composite struts
  is presented. It is adapted from an earlier successful study for
  confined layered structures which was formulated to model kink band
  formation in the folding of geological layers. The principal aim is
  to explore the underlying mechanisms governing the kinking response
  of flat, laminated components comprising unidirectional composite
  laminae. A pilot parametric study suggests that the key features of
  the mechanical response are captured well and that quantitative
  comparisons with experiments
  presented in the literature are highly encouraging.
\end{abstract}

\section{Introduction}

Kink banding is a phenomenon seen across many scales. It is a
potential failure mode for any layered, laminated or fibrous material,
held together by external pressure or some form of internal matrix,
and subjected to compression parallel to the layers. Many examples can
be found in the literature concerning the deformation of geological strata
\cite{Anderson,HoMeWi,PrCo}, wood and fibre composites
\cite{BudianskyFleck93,Kyriakides95,ReidPeng97,KyriakidesExpt,%
  Byskov2002,DaSilva20078685,Pimenta1_CST_2009}, and internally in
wire and fibre ropes \cite{Hobbs_textile,MAW_cst}. There have been
many attempts to reproduce kink banding theoretically, from early
mechanical models \cite{Rosen,Argon}, to more sophisticated
formulations derived from both continuum mechanics \cite{Budiansky83},
finite elasticity theory \cite{FuZhang06}
and numerical perspectives for more complex loading arrangements
\cite{KyriakidesFE}.

There has been much relevant work on composite materials with
significant problems being encountered as outlined
thus. First, although two dimensional models are commonly
employed \cite{BudianskyFleckAmazigo98,Pimenta2_CST_2009}, modelling
into the third dimension adds a significant extra component. It
inevitably involves a smeared approach in the modelling of material
properties since there is a mix of laminae and the matrix with the
possibility of voids. Secondly, failure is likely to be governed by
nonlinear material effects in shearing the matrix material
\cite{Fleck97}, and this is considerably less easy to measure or
control than the combination of overburden pressure and friction
considered in work on kink banding during geological folding
\cite{MAW_jsg,MAW_jmps04,MAW_jmps05}.

In the current paper, a pilot study is presented where the discrete
model formulated for kink banding in geological layers is adapted such
that it can be applied to unidirectional laminated composite struts
that are compressed in a direction parallel to the laminae.  This is
achieved by releasing the assumption that voids are wholly penalized
since, in the current case, no overburden pressure actively compresses
the layers in the transverse direction. Therefore, the rotation of the
laminae during the formation of the kink band causes a dilation, which
is resisted by transverse tensile forces generated within the
interlamina region. The coincident shearing of this region also
generates an additional resisting force, but, as mentioned above, this
can be subject to nonlinearity, in particular a reduced stiffness that
may be either positive (hardening) or negative (softening perhaps
leading to fracture), which is currently formulated with a piecewise
linear constitutive law. Work done from dilation and shearing are
evaluated; additional features from the original model: strain
energies from bending and direct compression, and the work done from
the external load can be incorporated without significant
alterations. An advantage of the presented model is that the resulting
equilibrium equations can be written and solved entirely in an
analytical form without having to resort to complex continuum models
or numerical solvers.

The primary aim of the current work is to lay the foundations for
future research. The geometric approach has yielded excellent
comparisons with experiments for the model for kink banding in
geological layers; the same is true currently with the present model
being compared favourably to previously published experiments
\cite{Kyriakides95}. Moreover, the relative importance of the
parameters governing the mechanical response are also identified in
the current study. From this, conclusions are drawn about the
possible further studies that would extend the current model to give
meaningful comparisons with the actual structural response for a
variety of practically significant scenarios.

\section{Review of model for geological layers}
\label{sec:geol_model}

\begin{figure}[htb]
  \centerline{\psfig{figure=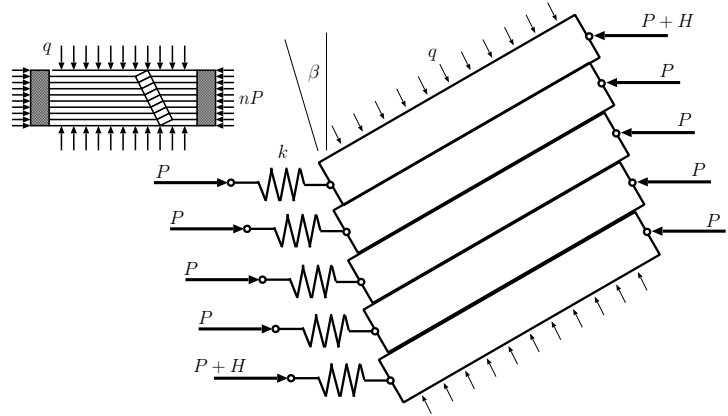,height=55mm}}
  \caption{Basic configuration of the discrete model for kink banding
    in $n$ geological layers. The tectonic load on each layer is $P$
    with a horizontal reaction force $H$, the axial stiffness of each
    layer is $k$, the overburden pressure on the layers is $q$ and the
    kink band orientation angle is $\beta$.}
  \label{fig:kinkgeol_overall}
\end{figure}

\begin{figure}[htb]
  \centerline{\psfig{figure=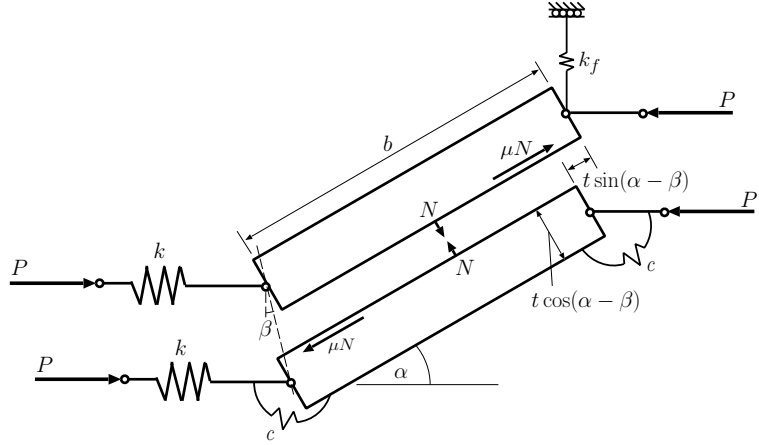,width=100mm}}
  \caption{Two internal layers of the geological model. The kink band
    width is $b$, the normal contact force between layers is $N$, the
    coefficient of friction is $\mu$, the stiffness of individual
    rotational springs modelling bending is $c$, the elastic stiffness
    of surrounding medium per unit layer is $k_f$ and the kink band
    angle is $\alpha$. A key assumption is the transverse
    compressibility of the layers.}
  \label{fig:kinkgeol_2layers}
\end{figure}

\begin{figure}[htb]
  \centerline{\psfig{figure=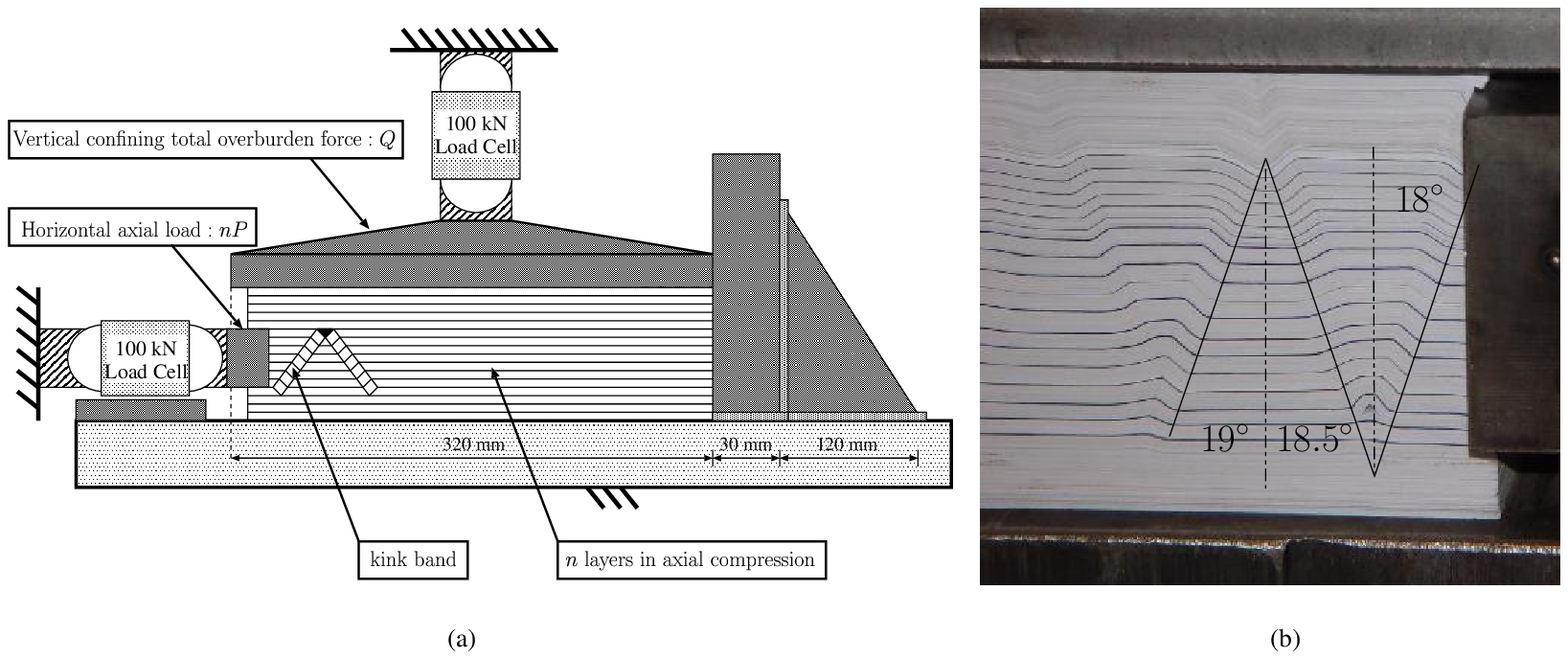,width=135mm}}
  \caption{(a) Schematic of the experimental rig used for testing the
    geological model. (b) A typical deformation profile in a physical
    experiment showing a sequence of kink bands with corresponding
    orientation angles $\beta$.}
  \label{fig:geol_rig}
\end{figure}

\begin{figure}[htb]
  \centerline{\psfig{figure=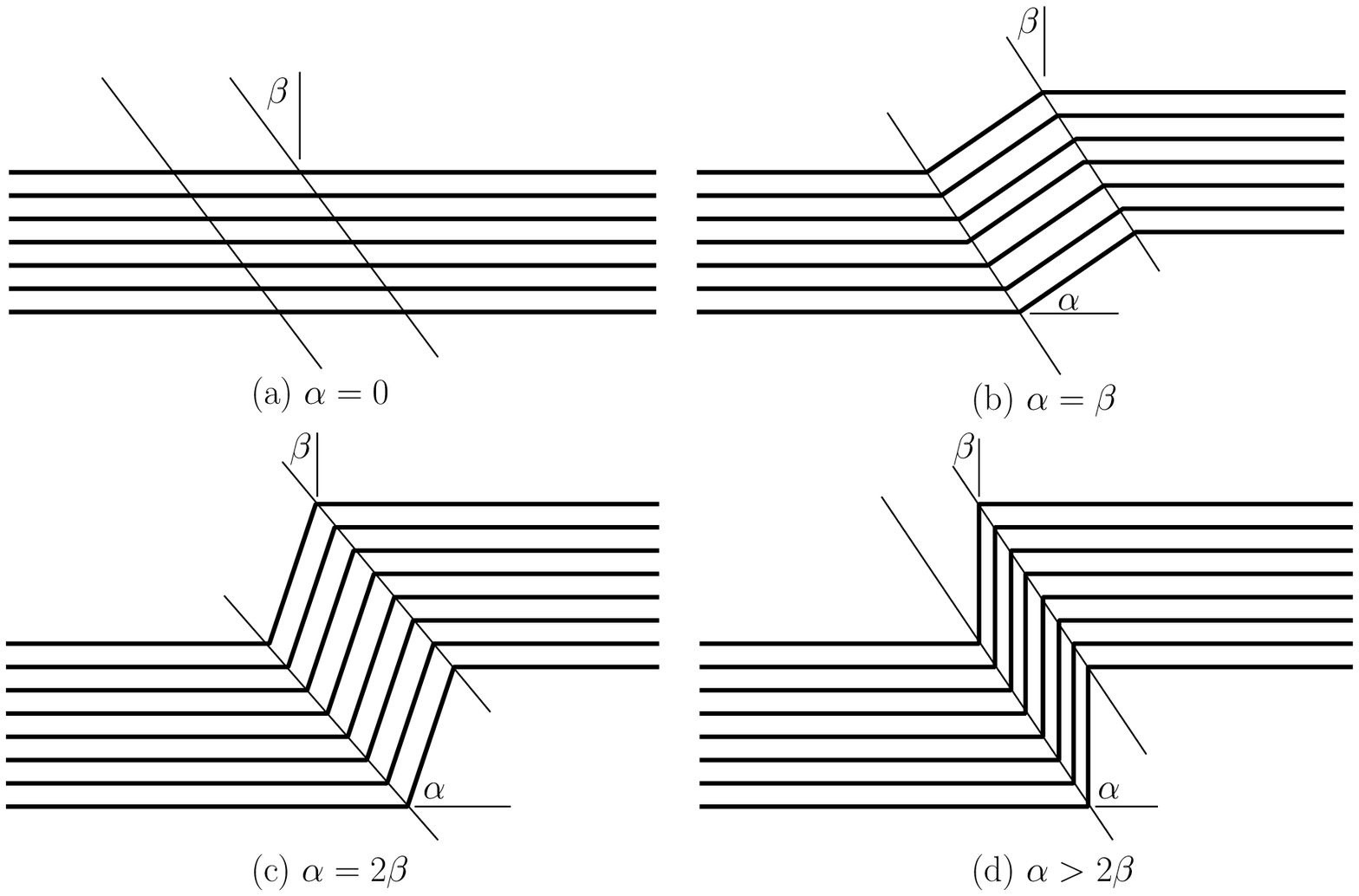,width=110mm}}
  \caption{Sequence of kink band deformation in the geological folding
    model. (a) initial state with $\beta$ defined from applied $q$;
    (b) instantaneous release of contact and hence friction within the
    kink band when $\alpha=\beta$; (c) layers inside and outside of
    the kink band all have equal thickness when $\alpha=2\beta$; (d)
    lock-up occurs when $\alpha>2\beta$ and a new band would form.}
  \label{fig:kinkgeol_seq}
\end{figure}

\begin{figure}[htb]
  \centerline{\psfig{figure=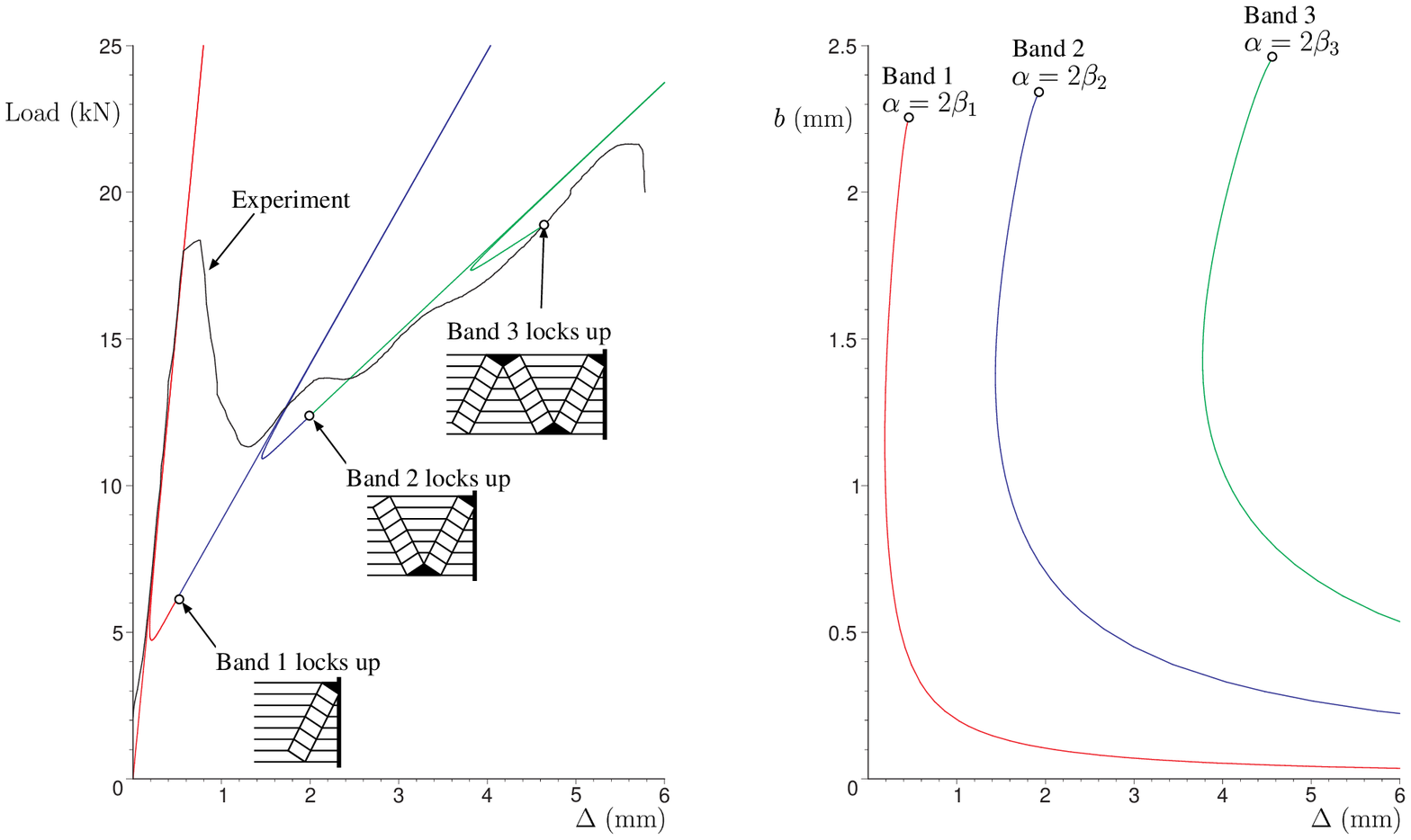,width=140mm}}
  \caption{Graphs showing the total axial load $P$ (left) and the kink
    band width $b$ (right) both versus the total end-displacement
    $\Delta$, taken from \protect\cite{MAW_jmps05}. The load versus
    displacement graph shows the lockup of the current kink band as a
    circle on the theoretical curve.}
  \label{fig:pbdel_expt4}
\end{figure}

A discrete formulation comprising springs, rigid links and Coulomb
friction has been devised to model kink band deformation in geological
layers that are held together by an overburden pressure
\cite{MAW_jmps04}. It was formulated using energy principles and key
parts of the model are shown in Figures \ref{fig:kinkgeol_overall}
and \ref{fig:kinkgeol_2layers}.
It has been compared very favourably with simple laboratory
experiments on layers of paper that were compressed transversely and
then increasingly compressed axially to trigger the kink band
formation process. The testing rig used in that study is shown
schematically in Figure \ref{fig:geol_rig}(a)
and a typical test photograph is shown in Figure
\ref{fig:geol_rig}(b). Assuming that the layers were transversely
compressible, the kink band orientation angle $\beta$ was predicted
theoretically for the first time; it being related purely to the
initially applied transverse strain derived from the overburden
pressure $q$. Figure \ref{fig:kinkgeol_seq}
shows the characteristic sequence of deformation with (a) showing the
undeformed state with applied overburden pressure and the transverse
pre-compression defining $\beta$; (b) showing the point where the
interlayer friction is released when the internal transverse strain
within the kink band is instantaneously zero and the band forming very
quickly in the direction of $\beta$. It was later demonstrated that
beyond the condition shown in Figure \ref{fig:kinkgeol_seq}(c), where
all the layers have the same thickness, whether internal or external to
the kink band, lock-up begins to occur as shown in Figure
\ref{fig:kinkgeol_seq}(d). This marked the point where new kink bands
formed and these could also be predicted by this approach after some
modifications were made to the model \cite{MAW_jmps05}. An example
comparison between the theory and an experiment from that study is
presented in Figure \ref{fig:pbdel_expt4}.
Moreover, this model has also been demonstrated to be suitable for
modelling internal kink band formation in individual composite fibres
under bending that are common in fibre ropes \cite{MAW_cst}.

\section{Pilot model for laminated composite struts}
\label{sec:compos_model}

As discussed above, the system studied in \cite{MAW_jmps04} had layers
that were bound together by the mechanisms of overburden pressure and
interlayer friction. The deformation was in fact only admissible
geometrically if the layers were transversely compressible; the
relationship between the kink band angle $\alpha$, which could vary,
and the orientation angle $\beta$, which was fixed, being such that
interlayer gaps, or voids, were not created. For a laminated strut,
most experimental evidence from the literature also suggests that the
kink band orientation angle $\beta$ is basically fixed for each
laminate configuration \cite{Kyriakides95,KyriakidesExpt}. It is
noted, however, in a recent study on laminates under combined
compression and shear that this angle can change as the kink
propagates but the angle reaches a limit \cite{Gutkin2_CST_2010}; in
the current work, $\beta$ is taken as a constant equivalent to this
limiting value from the beginning of the kink band deformation
process, which is a simplifying assumption.

\begin{figure}[htb]
  \centerline{\psfig{figure=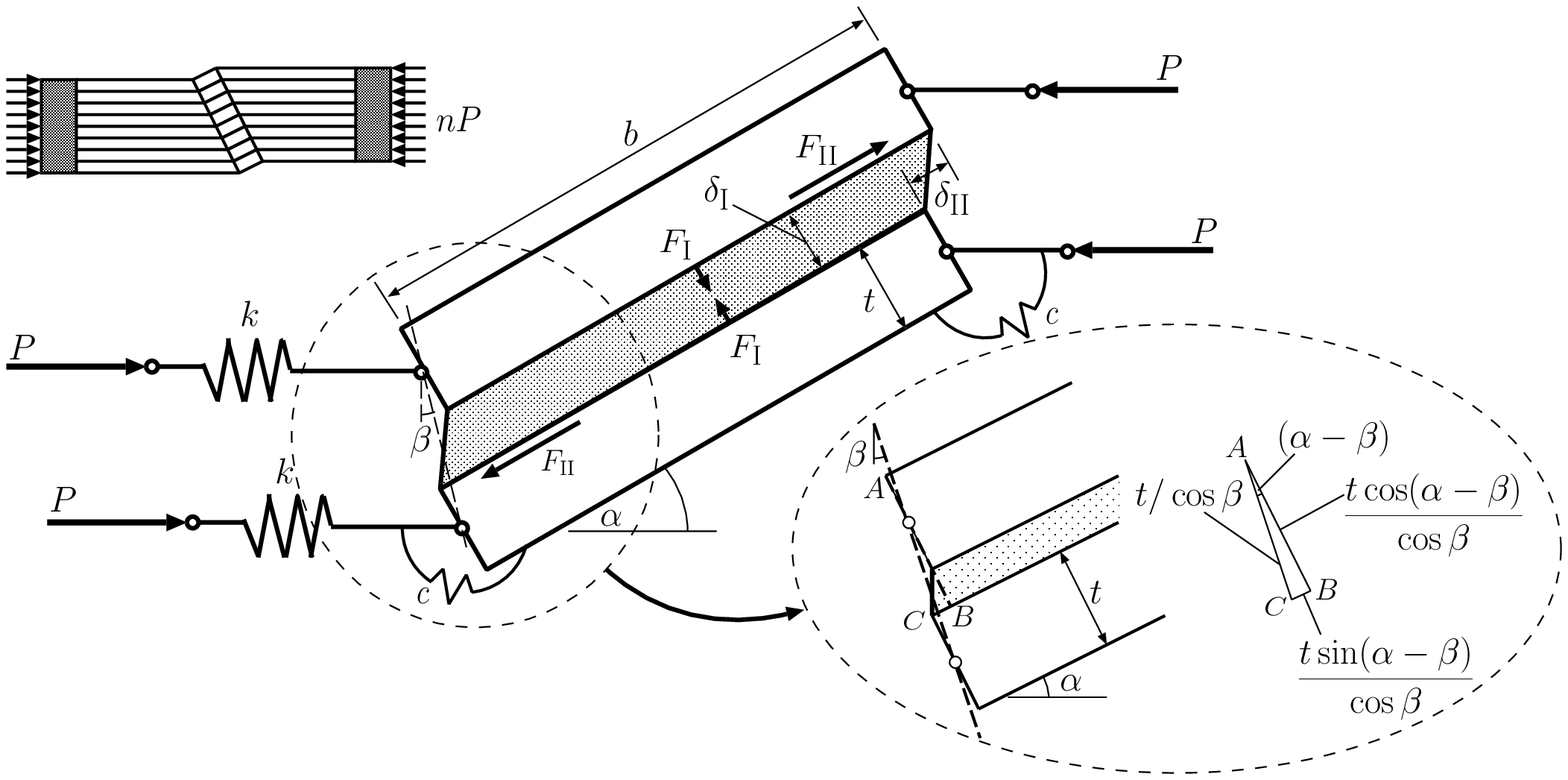,width=140mm}}
  \caption{Two internal laminae of the laminated composite model, the
    shaded region shows the interlamina region which is exaggerated in
    scale for clarity. Dilation and shearing forces with their
    corresponding displacements are given by: $F_\mathrm{I}$ and
    $F_\mathrm{II}$ with $\delta_\mathrm{I}$ and $\delta_\mathrm{II}$
    respectively. The highlighted section shows the lengths $AB$ and
    $BC$ that directly relate to $\delta_\mathrm{I}$ and
    $\delta_\mathrm{II}$ respectively.}
  \label{fig:compos_2layers}
\end{figure}

Kink band deformation in laminates involves different mechanisms that
incorporate the interlamina region comprising the laminae and the
matrix that binds the component together. Since the matrix is itself
deformable and that there is no overburden pressure to close any
voids, the model needs significant modifications to account for the
different characteristics of the laminated strut. It is worth noting
that the assumption for the lay-up sequence of the composite in the
present case is such that no twisting is generated from the applied
compression. Figure \ref{fig:compos_2layers}
shows the adapted 2-layer model which omits the following features
that are not relevant in the current case: the foundation stiffness
and the overburden pressure, \emph{i.e.}\ $q=k_f=0$. The kink band
formation is thus intrinsically linked to the deformation of the
interlamina region within the strut.

\begin{figure}[htb]
  \centerline{\psfig{figure=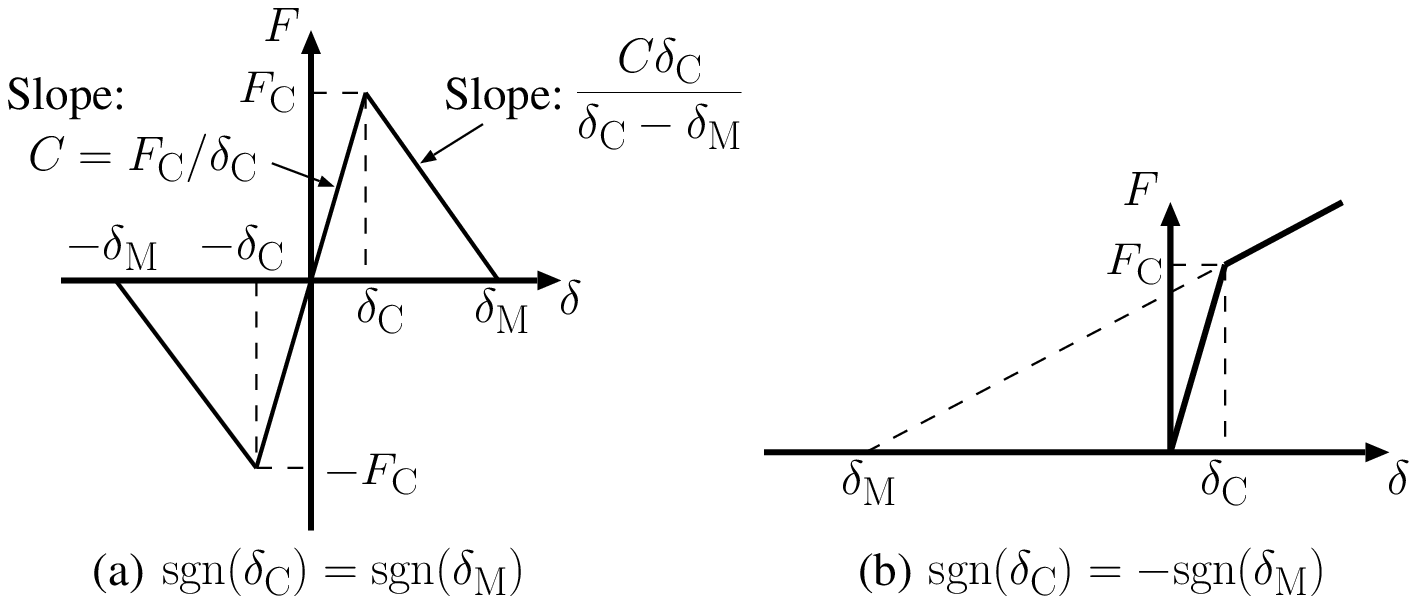,height=37mm}}
  \caption{Piecewise linear force versus displacement model applied
    for interlamina shearing: (a) a linear--softening response which is
    more representative of a fracture model; (b) a linear--hardening
    response which is more appropriate for materials that show
    post-yield strength.}
  \label{fig:fracturelaw}
\end{figure}

\begin{figure}[htb]
  \centerline{
    \subfigure[Mode I]{\psfig{figure=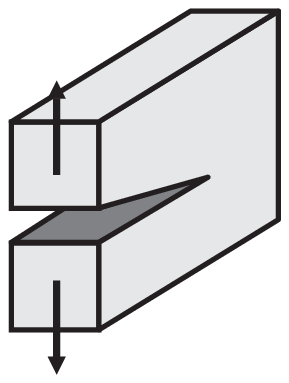,height=25mm}}\qquad
    \subfigure[Mode II]{\psfig{figure=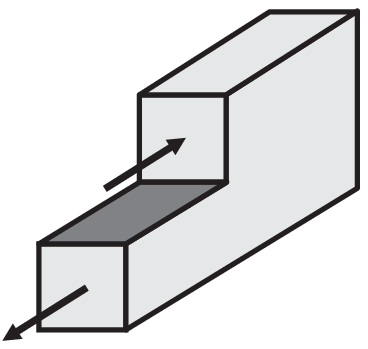,height=25mm}}\qquad
    \subfigure[Mode III]{\psfig{figure=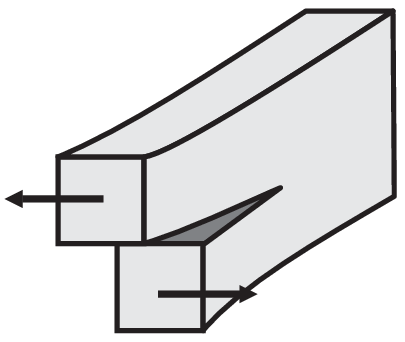,height=25mm}}}
  \caption{Fracture modes. Mode I is tearing; Mode II is shearing and
    Mode III is scissoring. In the current model, only Mode II is relevant.}
  \label{fig:fractmodes}
\end{figure}

Shearing within the interlamina region is the analogous process to
sliding between the layers in the model for geological folding; the
latter being modelled in the energy formulation as a work done
overcoming the friction force. A piecewise linear model is used to
simulate the force versus displacement relationship in terms of the
shear resistance (see Figure \ref{fig:fracturelaw}),
where fracture modes that are relevant for a linear--softening
response, see Figure \ref{fig:fracturelaw}(a), are defined as
in Figure \ref{fig:fractmodes}.

Tensile expansion, or dilation, of the interlamina region is modelled,
however, with a purely linear elastic constitutive law. In the model
described in \S\ref{sec:geol_model}, it was argued that when the
interlayer contact force was released that not only would the friction
be released but also that the overburden pressure would inhibit the
formation of subsequent voids within the layered structure. Since in
the current case there is no overburden or lateral pressure as such,
potential dilation of the interlamina region needs to be included. As
in the previous model, however, the lamina deformation is assumed to
lock-up and potentially trigger a new band forming when $\alpha >
2\beta$; transverse compression in adjacent laminae would then be
occurring and stiffening the response significantly. Hence, it is
reasoned therefore that it would be energetically advantageous for the
mechanical system to form a new kink band rather than continue to
deform the current one \cite{MAW_jmps05}.

\subsection{Potential energy formulation}

\subsubsection{Interlamina dilation}

The resistance to interlamina dilation while the kink bands deform is
modelled with a linear constitutive law; the dilation resisting force
$F_\mathrm{I}$ relating to the dilation displacement
$\delta_\mathrm{I}$, thus:
\begin{equation}
  F_\mathrm{I} (\alpha) = C_\mathrm{I} \delta_\mathrm{I}, \quad
  \delta_\mathrm{I} (\alpha) = t \left[ \frac{\cos (\alpha - \beta)}{\cos
      \beta} -1 \right],
  \label{eq:delta_I}
\end{equation}
with $C_\mathrm{I}$ being the transverse stiffness of the laminate,
related to the transverse Young's modulus, and $t$ being the thickness
of an individual lamina. Since the area over which the interlamina
region dilates depends directly on the kink band width $b$, the
stiffness $C_\mathrm{I}$ can be expressed as:
\begin{equation}
  C_\mathrm{I} = b d k_\mathrm{I},
\end{equation}
where $k_\mathrm{I}$ is the transverse stiffness per unit area of the
laminate and $d$ is the breadth of the strut. However, with the lamina
assumed to be transversely incompressible in the current model and the
dilation displacement being assigned purely to the softer interlamina
matrix material, a clear departure from the geological model, the
current lamina thickness is thus $t$ rather than
$t\cos(\alpha-\beta)$. This is shown in Figure
\ref{fig:compos_2layers} and is detailed in the highlighted area of
that diagram. The relationship in equation (\ref{eq:delta_I}) for
$\delta_\mathrm{I}$ is thus obtained from taking the length $AB$ from
Figure \ref{fig:compos_2layers}, where
$\delta_\mathrm{I}=AB-t$. Hence, there is a transverse tensile strain
developed as the gap between the laminae grows as $\alpha$ increases
from zero to $\beta$. The gap subsequently begins to reduce; when
$\alpha=2\beta$ the gap returns to zero, marking the commencement of
lock-up.

The work done in the dilation process is therefore given by the
expression:
\begin{equation}
  U_D = \int_0^{\delta_\mathrm{I} (\alpha)} F_\mathrm{I}(\alpha')
  \,\mathrm{d}\left\{t \left[ \frac{\cos (\alpha' - \beta)}{\cos
        \beta} -1 \right] \right\} = \frac{k_\mathrm{I} bdt^2}{2}
  \left[ 1 - \frac{\cos(\alpha - \beta)}{\cos \beta} \right]^2.
  \label{eq:UD_linear}
\end{equation}
It is assumed that the interlamina region would not be damaged in the
process of dilation and that the only nonlinearity in the constitutive
law would be under shear. This is because the dilation displacement is
relatively smaller than the shearing displacement that is discussed
next; this has the additional advantage of maintaining model
simplicity such that any mixed mode fracture considerations can be
left for future work.

\subsubsection{Interlamina shearing}

Interlamina shearing or the laminae sliding relative to one another is
modelled with a piecewise linear constitutive law with the force
resisting shear $F_\mathrm{II}$ relating to the shearing displacement
$\delta_\mathrm{II}$, thus:
\begin{equation}
  F_\mathrm{II} (\alpha) = C_\mathrm{II} \delta_\mathrm{II}, \quad
  \delta_\mathrm{II} (\alpha) = \frac{t}{\cos\beta} \left[ \sin (\alpha-\beta)
  + \sin \beta \right],
  \label{eq:delta_II}
\end{equation}
with $C_\mathrm{II}$ being the shearing stiffness of the combination
of the matrix and laminae sliding relative to one another. The
relationship for $\delta_\mathrm{II}$ in terms of $\alpha$ and $\beta$
in equation (\ref{eq:delta_II}) is given by examining the length $BC$
in Figure \ref{fig:compos_2layers}. However, since the band is
basically assumed to form instantaneously before any rotation occurs,
it is implied that
$F_\mathrm{II}(0)=\delta_\mathrm{II}(0)=0$. However, since $\beta \neq
0$, the expression $\delta_\mathrm{II}=BC+t \tan\beta$ is obtained,
where the force and displacement conditions are satisfied.

When the shearing displacement reaches the initial proportionality
limit, \emph{i.e.}\ when $\delta_\mathrm{II} = \delta_\mathrm{C}$ (see
Figure \ref{fig:fracturelaw}), the relationship between
$F_\mathrm{II}$ and $\delta_\mathrm{II}$ changes to:
\begin{equation}
  F_\mathrm{II} = C_\mathrm{II}  \delta_\mathrm{C} \left( 
    \frac{\delta_\mathrm{II} - \delta_\mathrm{M}}{
      \delta_\mathrm{C} - \delta_\mathrm{M}} \right),
\end{equation}
where $\delta_\mathrm{M}$ is the shearing displacement when the
corresponding resistance force reduces to zero. Now, if
$\delta_\mathrm{II}(\alpha_\mathrm{C}) = \delta_\mathrm{C}$ and
$\delta_\mathrm{II} (\alpha_\mathrm{M}) = \delta_\mathrm{M}$, the
expressions for the resisting force can be written thus:
\begin{equation}
  F_\mathrm{II} =
  \begin{cases}
    & C_\mathrm{II} t \left[ \sin (\alpha-\beta) + \sin\beta \right] /
    \cos\beta \quad \text{for }\alpha=\left[
      0,\alpha_\mathrm{C} \right], \\
    & \cfrac{C_\mathrm{II} t}{\cos\beta} \left[ \cfrac{ \sin (\alpha -
        \beta) - \sin (\alpha_\mathrm{M} - \beta)}{%
        \sin (\alpha_\mathrm{C} - \beta) - \sin (\alpha_\mathrm{M} -
        \beta)} \right] \left[ \sin (\alpha_\mathrm{C}-\beta)
      + \sin\beta \right] \quad\text{for } \alpha > \alpha_\mathrm{C}\\
    & \qquad \qquad \qquad \qquad \qquad \qquad \qquad \qquad \qquad
    \text{and }\alpha = \left[ \alpha_\mathrm{C},\alpha_\mathrm{M}
    \right] \text{ if }\sgn(\delta_\mathrm{C}) = \sgn(\delta_\mathrm{M})>0, \\
    & 0 \quad \text{for }\alpha \geqslant \alpha_\mathrm{M}\text{ and }
    \sgn(\delta_\mathrm{C}) = \sgn(\delta_\mathrm{M})>0.
  \end{cases}  
\end{equation}
Moreover, since the shear area of contact depends on the kink band
width $b$, the stiffness $C_\mathrm{II}$ can be expressed as:
\begin{equation}
  C_\mathrm{II} = b d k_\mathrm{II},
\end{equation}
where $k_\mathrm{II}$ is the shear stiffness per unit area of the
lamina and hence the effective shear stress $\tau$ is defined:
\begin{equation}
  \label{eq:tau}
  \tau = k_\mathrm{II} \delta_\mathrm{II}.
\end{equation}
The work done in the shearing process is given by the expression:
\begin{equation}
  \begin{split}
  U_S &= \int_{0}^{\delta_\mathrm{II}(\alpha)} F_\mathrm{II}(\alpha') \,\mathrm{d}
  \left\{ \frac{t}{\cos\beta}  \left[ \sin (\alpha'-\beta) +
      \sin\beta \right] \right\}\\
  &= \frac{k_\mathrm{II} bdt^2}{2}
  \left[ \frac{\sin (\alpha-\beta) + \sin\beta}{\cos\beta} \right]^2\\
  &= \frac{k_\mathrm{II} bdt^2}{2} \mathcal{L}(\alpha),
  \end{split}
\end{equation}
for $\alpha \leqslant \alpha_\mathrm{C}$, or:
\begin{equation}
  \begin{split}    
    U_S &= \frac{k_\mathrm{II} bdt^2}{2} \biggl\{
    \mathcal{L}(\alpha_\mathrm{C})\\
    & \qquad + \int_{\alpha_\mathrm{C}}^\alpha
    \left[ \frac{ \sin (\alpha' - \beta) - \sin (\alpha_\mathrm{M} -
        \beta)}{%
        \sin (\alpha_\mathrm{C} - \beta) - \sin (\alpha_\mathrm{M} -
        \beta)} \right] \left[ \frac{\sin (\alpha_\mathrm{C}-\beta) +
        \sin\beta}{\cos^2\beta} \right] \cos(\alpha'-\beta) \,
    \mathrm{d}\alpha' \biggr\} \\
    &= \frac{k_\mathrm{II} bdt^2}{2 \cos^2\beta} \biggl\{ \left[
      \sin(\alpha_\mathrm{C}-\beta) + \sin\beta \right]^2 + \left[
      \frac{ \sin(\alpha_\mathrm{C}-\beta) + \sin\beta}{%
        \sin(\alpha_\mathrm{C}-\beta) - \sin(\alpha_\mathrm{M}-\beta)}
    \right] \biggl[
    \sin^2 (\alpha-\beta) \\
    & \qquad\qquad\qquad - \sin^2 (\alpha_\mathrm{C}-\beta) + 2
    \sin(\alpha_\mathrm{M}-\beta) [ \sin (\alpha_\mathrm{C}-\beta) -
    \sin(\alpha-\beta) ]
    \biggr] \biggr\} \\
    &= \frac{k_\mathrm{II} bdt^2}{2} \mathcal{S}(\alpha),
  \end{split}
\end{equation}
beyond the proportionality limit where $\alpha=\left[
  \alpha_\mathrm{C},\alpha_\mathrm{M} \right]$. However, if
$\alpha>\alpha_\mathrm{M}$ and $\sgn(\delta_\mathrm{C}) =
\sgn(\delta_\mathrm{M})>0$ the shear resistance force vanishes and the
expression for $U_S$ becomes:
\begin{equation}
  \begin{split}    
    U_S &= \frac{k_\mathrm{II} bdt^2}{2} \biggl\{
    \mathcal{L}(\alpha_\mathrm{C}) \\
    & \qquad +
    \int_{\alpha_\mathrm{C}}^{\alpha_\mathrm{M}} \left[ \frac{ \sin
        (\alpha' - \beta) - \sin (\alpha_\mathrm{M} - \beta)}{%
        \sin (\alpha_\mathrm{C} - \beta) - \sin (\alpha_\mathrm{M} -
        \beta)} \right] \left[ \frac{\sin (\alpha_\mathrm{C}-\beta) +
        \sin\beta}{\cos^2\beta} \right] \cos(\alpha'-\beta) \,
    \mathrm{d}\alpha' \biggr\} \\
    &= \frac{k_\mathrm{II} bdt^2}{2 \cos^2\beta} \biggl\{ \sin\beta
    \left[ \sin\beta + \sin (\alpha_\mathrm{C}-\beta) + \sin
      (\alpha_\mathrm{M}-\beta) \right]
    + \sin (\alpha_\mathrm{C}-\beta) \sin (\alpha_\mathrm{M}-\beta) \biggr\} \\
    &= \frac{k_\mathrm{II} bdt^2}{2} \mathcal{S}(\alpha_\mathrm{M}).
  \end{split}
\end{equation}
There would still be the potential for frictional forces to resist
shear even though $\delta>\delta_\mathrm{M}$ and the interlamina
region has lost all shear strength. However, this effect is currently
neglected and hence the current model would tend to underestimate the
true strength to some extent.

\subsubsection{Remaining energy contributions}

\begin{figure}[htb]
  \centerline{\psfig{figure=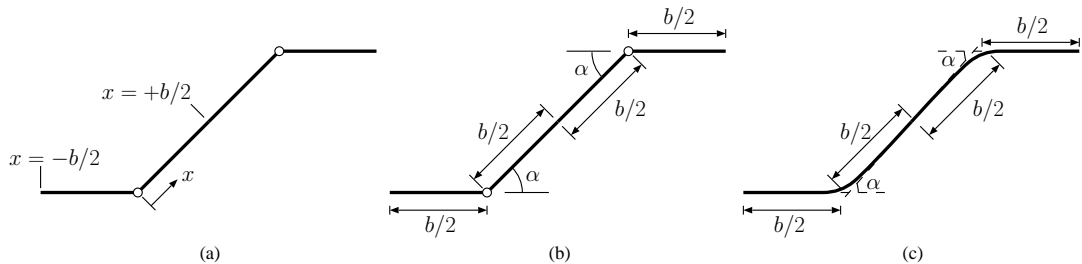,height=35mm}}
  \caption{Bending of a lamina: (a) definition of $x$; (b) idealized
    case; (b) actual case. Curvature $\kappa$ is defined as the total
    angle change $2\alpha$ over the effective length of the band $2b$,
    hence $\kappa \approx \alpha/b$.}
  \label{fig:bendc}
\end{figure}

As in the geological model, the strain energy stored in
bending can be taken from a pair of rotational springs of stiffness $c$:
\begin{equation}
  \label{eq:Ub}
  U_b = c \alpha^2.
\end{equation}
The stiffness of the rotational springs is related differently from
the geological model as the expression for that model contained the
overburden pressure $q$ \cite{MAW_jmps04}. Since the bending energy
should strictly relate to curvature $\kappa$, where:
\begin{equation}
  U_b = 2\int_{-b/2}^{b/2} \frac12 EI \kappa^2 \, \mathrm{d}x,
\end{equation}
with $x$ defining the domain of one bending corner and $\kappa$ as the
rate of change of the kink band angle $\alpha$ over the kink band
width $b$, as represented in Figure \ref{fig:bendc},
thus: 
\begin{equation}
  \kappa \approx \frac{\alpha}{b} \Rightarrow c \approx \frac{EI}{b}.
\end{equation}
Hence, the rotational stiffness $c$ is related to the flexural
rigidity $EI$ of a lamina with $E$ being its the Young's modulus in
the axial direction and its second moment of area $I=dt^3/12$.  The
strain energy per layer associated with the in-line spring of
stiffness $k$ is hence given by:
\begin{equation}
  \label{eq:Um}
  U_m = \frac12 k \delta_a^2,
\end{equation}
where $\delta_a$ is the axial displacement of the springs. The in-line
spring stiffness $k=Edt/L$ for a single lamina with $L$
being the length of the strut.  The work done by the external load can
be taken simply as the sum of the displacement of the in-line springs
$\delta_a$ and from the band deforming multiplied by the axial load
$P$, which can be defined as the axial pressure $p$ multiplied by the
cross-sectional area of a lamina, $dt$:
\begin{equation}
  \label{eq:PDel}
  P\Delta = pdt \left[ \delta_a + b (1-\cos\alpha) \right].
\end{equation}

\subsubsection{Total potential energy functions}

The total potential energy $V$ is given by the sum of the strain
energies from bending $U_b$, the in-line springs $U_m$, interlaminar
dilation $U_D$ and shearing $U_S$, minus the work done $P\Delta$,
thus:
\begin{equation}
  \label{eq:TPE}
  V = U_b + U_m + U_D + U_S - P\Delta.
\end{equation}
Since the dilation terms are assumed to be linearly elastic throughout
their loading history, the total potential energy per axially loaded
lamina takes three forms:
\begin{enumerate}
\item The case where $\alpha = [0,\alpha_\mathrm{C}]$, so
  $V=V^\mathrm{L}$, \emph{i.e.} linearly elastic in shear.
\item The case where $\alpha > \alpha_\mathrm{C}$, so $V=V^\mathrm{S}$,
  \emph{i.e.} the secondary shear stiffness is either a smaller positive
  value than the primary shear stiffness or a negative value.
\item The case where $\alpha > \alpha_\mathrm{M}$ and
  $\sgn(\delta_\mathrm{C}) = \sgn(\delta_\mathrm{M})>0$, so
  $V=V^\mathrm{Z}$, \emph{i.e.} no shear stiffness, which only occurs
  if the secondary shear stiffness is negative.
\end{enumerate}
These forms of the total potential energy are given by the expressions:
\begin{equation}
  \label{eq:VL_VS}
  V^\mathrm{L} = V^\mathrm{I} + \frac{k_\mathrm{II} bdt^2}{2}
  \mathcal{L}(\alpha), \quad
    V^\mathrm{S} = V^\mathrm{I} + \frac{k_\mathrm{II} bdt^2}{2}
    \mathcal{S}(\alpha), \quad
  V^\mathrm{Z} = V^\mathrm{I} + \frac{k_\mathrm{II} bdt^2}{2}
    \mathcal{S}(\alpha_\mathrm{M}),
\end{equation}
where $V^\mathrm{I}$ is given by:
\begin{equation}
  \label{eq:VI}
    V^\mathrm{I} = \frac{k\delta_a^2}{2}  + \frac{Edt^3\alpha^2}{12b}
    + \frac{k_\mathrm{I} bdt^2}{2} \left[ 1 -
      \frac{\cos(\alpha - \beta)}{\cos \beta} \right]^2 - pdt \left[
      \delta_a + b (1-\cos\alpha) \right].
\end{equation}
The total potential energy functions are nondimensionalized by
dividing through by $kt^2$ and can be re-expressed in terms of
rescaled parameters:
\begin{equation}
  \label{eq:V_nondim}
  \tilde{V}^\mathrm{L} = \tilde{V}^\mathrm{I} +
  \frac{\tilde{k}_\mathrm{II} \tilde{b}}{2} \mathcal{L}(\alpha), \quad
  \tilde{V}^\mathrm{S} = \tilde{V}^\mathrm{I} +
  \frac{\tilde{k}_\mathrm{II}\tilde{b}}{2} \mathcal{S} (\alpha), \quad
  \tilde{V}^\mathrm{Z} = \tilde{V}^\mathrm{I} +
  \frac{\tilde{k}_\mathrm{II}\tilde{b}}{2} \mathcal{S}
  (\alpha_\mathrm{M}),
\end{equation}
where:
\begin{equation}
  \begin{split}
    \label{eq:nondim}
    \tilde{V}^\mathrm{I} &= \frac{V^\mathrm{I}}{kt^2},\quad
    \tilde{V}^\mathrm{L} = \frac{V^\mathrm{L}}{kt^2},\quad
    \tilde{V}^\mathrm{S} = \frac{V^\mathrm{S}}{kt^2},\quad
    \tilde{V}^\mathrm{Z} = \frac{V^\mathrm{Z}}{kt^2},\quad
    \tilde{\delta}=\frac{\delta_a}{t}, \quad
    \tilde{\Delta}=\frac{\Delta}{t},\\
    \tilde{b} = \frac{b}{t}, \quad
    \tilde{p} &= \frac{pd}{k} = \frac{pL}{Et},\quad 
    \tilde{D} = \frac{Ed}{12k} = \frac{L}{12t},\quad
    \tilde{k}_\mathrm{I} = \frac{k_\mathrm{I}dt}{k} =
    \frac{k_\mathrm{I}L}{E}, \quad
    \tilde{k}_\mathrm{II} = \frac{k_\mathrm{II}dt}{k} = \frac{k_\mathrm{II}L}{E}.
  \end{split}
\end{equation}

\subsection{Equilibrium equations}

The equilibrium equations are defined by the condition of stationary
potential energy with respect to the end-shortening $\delta_a$, the
kink band angle $\alpha$ and the kink band width $b$; these can be
written in nondimensional terms, thus:
\begin{align}
  \label{eq:equil_L_delta}
  \tilde{p} &= \tilde{\delta},\\
  \label{eq:equil_L_alpha}
  \tilde{p} &= \tilde{k}_{\mathrm{I}} I_\alpha + \tilde{k}_{\mathrm{II}} J_\alpha +
  \frac{2\tilde{D} \alpha}{\tilde{b}^2 \sin \alpha},\\
  \label{eq:equil_L_b}
  \tilde{p} &= \tilde{k}_{\mathrm{I}} I_b + \tilde{k}_{\mathrm{II}}
  J_b - \frac{\tilde{D}\alpha^2}{\tilde{b}^2\left( 1 - \cos \alpha \right)}.
\end{align}
Equation (\ref{eq:equil_L_delta}) defines the pre-kinking fundamental
equilibrium path that accounts for pure compression of the in-line
springs of stiffness $k$. Equations
(\ref{eq:equil_L_alpha})--(\ref{eq:equil_L_b}) define the
post-instability states for the non-trivial kink band deformations;
equating them allows the kink band width $b$ to be evaluated
analytically:
\begin{equation}
  \label{eq:bandwidth}
  \tilde{b} = \left\{\frac{\tilde{D} \alpha \left[2/\sin\alpha +
        \alpha/\left( 1 - \cos\alpha \right) \right]}{\tilde{k}_{\mathrm{I}}
      \left( I_b - I_\alpha \right) 
    + \tilde{k}_{\mathrm{II}} \left( J_b - J_\alpha \right)} \right\}^{1/2}.
\end{equation}
The expressions for $I_\alpha$ and $I_b$ are given in detail thus:
\begin{equation}
  \label{eq:Ia_Ib}
  I_\alpha = \left[1 - \frac{\cos(\alpha - \beta)}{\cos \beta} \right]
  \frac{\sin(\alpha-\beta)}{\sin \alpha \cos\beta}, \quad
  I_b =  \frac{1}{2\left( 1 - \cos \alpha \right)}
  \left[ 1 - \frac{\cos(\alpha - \beta)}{\cos \beta} \right]^2,
\end{equation}
where these expressions apply for the entire range of
$\alpha$. However, the expressions for $J_\alpha$ and $J_b$ change for
each form of the total potential energy function; for
$\tilde{V}=\tilde{V}^\mathrm{L}$, the expressions are:
\begin{equation}
  \label{eq:VL_J}
  J_\alpha =
  \frac{\cos(\alpha-\beta)\left[\sin(\alpha-\beta)+\sin\beta\right]}{\sin\alpha
    \cos^2\beta}, \quad 
  J_b = \frac{\left[\sin (\alpha-\beta) + \sin\beta
    \right]^2}{2(1-\cos\alpha)\cos^2\beta};
\end{equation}
for $\tilde{V}=\tilde{V}^\mathrm{S}$:
\begin{equation}
  \label{eq:VS_J}
  \begin{split}
    J_\alpha &= \frac{\cos(\alpha-\beta)}{\sin\alpha \cos^2\beta}
    \left[ \frac{\sin(\alpha_\mathrm{C}-\beta) + \sin\beta}{%
        \sin(\alpha_\mathrm{C}-\beta)-\sin(\alpha_\mathrm{M}-\beta)}
    \right] \left[ \sin(\alpha-\beta)-\sin(\alpha_\mathrm{M}-\beta) \right], 
    \\
    J_b &= \left[ \frac{\sin(\alpha_\mathrm{C}-\beta) + \sin\beta}{2(1-\cos
      \alpha) \cos^2 \beta} \right] \biggl\{ \sin(\alpha_\mathrm{C}-\beta) +
      \sin \beta \\
      & \qquad
      + \frac{ \sin^2 (\alpha-\beta) - \sin^2
        (\alpha_\mathrm{C}-\beta) + 2\sin(\alpha_\mathrm{M}-\beta)[
        \sin(\alpha_\mathrm{C}-\beta) -
        \sin(\alpha-\beta)]}{\sin(\alpha_\mathrm{C}-\beta) -
        \sin(\alpha_\mathrm{M}-\beta)} \biggr\};
  \end{split}
\end{equation}
and for $\tilde{V}=\tilde{V}^\mathrm{Z}$:
\begin{equation}
  \label{eq:VZ_J}
  \begin{split}
  J_\alpha = 0, \quad
  J_b = \frac{\sin\beta \left[ \sin\beta + \sin
      (\alpha_\mathrm{C}-\beta) + \sin (\alpha_\mathrm{M}-\beta)
    \right] + \sin (\alpha_\mathrm{C}-\beta) \sin
    (\alpha_\mathrm{M}-\beta)}{%
    2\cos^2\beta(1-\cos\alpha)}.
  \end{split}
\end{equation}
The initial limiting case where $\alpha \rightarrow 0$ gives $b
\rightarrow \infty$ and $\tilde{p} \rightarrow \tilde{k}_{\mathrm{I}}
\tan^2 \beta + \tilde{k}_{\mathrm{II}}$. The result for $b$ suggesting
that the kink band is initially prevalent throughout the structure and
the result for $p$ showing that the critical load depends primarily on
the shear stiffness with a smaller contribution from the dilation
stiffness that relates to $\beta$. This reproduces similar results
from the literature where the critical stress is related to the shear
modulus \cite{BudianskyFleck93,Kyriakides95,Rosen}; it also reflects a
significant difference from the geological model which has an infinite
critical load and where the kink band width grows from zero length
\cite{MAW_jmps04}.

\section{Numerical investigations}

\subsection{Validation against experimental results}
\label{sec:validation}

Results from the current model are initially compared with published
experiments on circular cylindrical composite rods with confined ends
that exhibited kink bands under axial compression
\cite{Kyriakides95}. Although the model is formulated for flat
rectangular laminae, the lamina thickness $t$ can be perceived to be
equivalent to the diameter of an individual fibre rod. This aids the
comparison between the current model and the experiments such that
both loading levels and the kink band width can be compared; a similar
approach was employed in \cite{MAW_cst}.

\begin{figure}[htb]
\centering
\psfig{figure=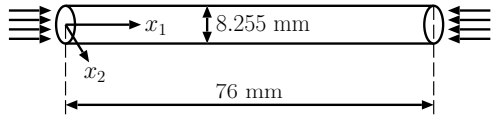,width=65mm}
\caption{Representation of the experimental sample in
  \protect\cite{Kyriakides95}. The rod comprised ICI APC-2/AS4
  composite fibres with properties as given in Table
  \protect\ref{tab:prop_Kyriakides}. The sample was confined such that
  there was negligible transverse compression but also that global
  buckling was not an issue.}
  \label{fig:rodexpt}
\end{figure}

The dimensions of the overall sample had a diameter of $8.255~\mm$
with the relevant properties given in Table
\ref{tab:prop_Kyriakides}.
\begin{table}
  \begin{tabbing}
    ***********************************\=\kill
    Rod fibre diameter: \> $t=7\times 10^{-3}~\mm$\\
    Overall rod length \> $L=76~\mm$\\
    Longitudinal Young's modulus: \> $E=E_{11} = 130.76~\mathrm{kN/mm^2}$\\
    Transverse Young's modulus: \> $E_{22} = 10.40~\mathrm{kN/mm^2}$\\
    Shear modulus (initial to final): \> $G_{12} = 6.03
    \rightarrow 0.68~\mathrm{kN/mm^2}$ over $4\%$ shear strain\\
    Nondimensional stiffness quantities:\\
    Flexural rigidity: \> $\tilde{D}=L/(12t)$ \\
    Dilation stiffness: \> $\tilde{k}_\mathrm{I}=E_{22}L/(E_{11}t)$\\
    Shearing stiffness: \> $\tilde{k}_\mathrm{II}=G_{12}L/(E_{11}t)$
  \end{tabbing}
  \caption{Properties used in the validation study to compare the
    current model with experiments presented in \protect\cite{Kyriakides95}.}
  \label{tab:prop_Kyriakides}
\end{table}
Note that the breadth $d$ is not given since it cancels in all the
relevant nondimensional quantities. The sample comprised ICI APC-2/AS4
composite fibres. Since the sample was cylindrical, the system in
\cite{Kyriakides95} was presented in terms of a cylindrical polar
coordinate system with $x_1$ and $x_2$ being the longitudinal and the
radial coordinates respectively, as shown in Figure \ref{fig:rodexpt}.
For the tests presented to measure the change in shear modulus
$G_{12}$, there was no plateau shown in the test data, see Figures A3
and A5 in \cite{Kyriakides95}. In the current study, it is therefore
assumed that the piecewise linear model for the shear stiffness
reflect the initial and final values found in the experiments; hence
$\sgn(\delta_\mathrm{M}) = - \sgn(\delta_\mathrm{C})$, \emph{i.e.}\ a
linear--hardening model is implemented as represented in Figure
\ref{fig:fracturelaw}(b).

The critical shear angle, $\gamma_\mathrm{C}$, which is effectively
equal to the so-called engineering shear strain, for the piecewise
linear idealization, is the angle beyond which the shear stiffness is
replaced by a secondary smaller value; this is estimated from the
aforementioned graphs in Figures A3 and A5 in \cite{Kyriakides95} to
be $0.012~\mathrm{rad}$ (or $0.69^\circ$). The shear angle can be
expressed in terms of the kink band orientation angle $\beta$ and the
kink band angle, $\alpha$, such that:
\begin{equation}
  \tan \gamma_\mathrm{C} =\frac{
    \delta_\mathrm{II}(\alpha_\mathrm{C})}{%
    \delta_\mathrm{I}(\alpha_\mathrm{C}) + t }
  =\frac{ \sin \left( \alpha_\mathrm{C} - \beta \right)
    + \sin \beta }{ \cos \left( \alpha_\mathrm{C} -\beta \right)}.
  \label{eq:gamC}
\end{equation}
Given that $\beta$ is assumed to remain constant during deformation,
the critical kink band angle $\alpha_\mathrm{C}$ can be found by
rearranging (\ref{eq:gamC}), thus:
\begin{equation}
  \tan \gamma_\mathrm{C} \cos \left(\alpha_\mathrm{C} - \beta \right)
  - \sin \left(\alpha_\mathrm{C} - \beta \right) -\sin \beta =0,
  \label{eq:gamC2}
\end{equation}
and solving for $\alpha_\mathrm{C}$.  This is achieved by substituting
the critical shear angle $\gamma_\mathrm{C}$ from above and the kink
band orientation angle from \cite{Kyriakides95}, where $\beta$ was
reported to lie between $12^{\circ} \rightarrow 16^{\circ}$ to the
$x_2$ direction; for the specified values, $\alpha_\mathrm{C}$ is
approximately equal to $\gamma_\mathrm{C}$ ($\alpha_\mathrm{C} =
0.0120 \rightarrow 0.0121~\mathrm{rad}$). To obtain the correct final
shear modulus (see Table \ref{tab:prop_Kyriakides}) the value of
$\delta_\mathrm{M}/t=-0.094$ is used such that the ratios between the
initial and final values of the shear stiffness reflect the reported
experimental data.

\begin{figure}[htb] \centering
\psfig{figure=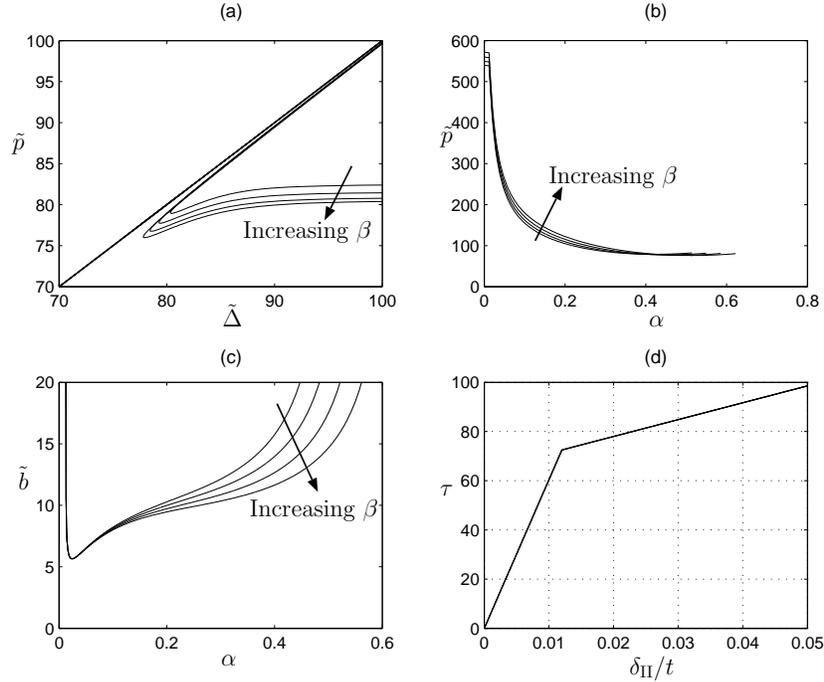,width=110mm}
\caption{Nondimensional plots of load $\tilde{p}$ versus (a) total
  end-shortening $\tilde{\Delta}$ and (b) kink band angle
  $\alpha~(\mathrm{rad})$ ; (c) kink band width $\tilde{b}$ versus
  kink band angle $\alpha~(\mathrm{rad})$. Range of $\beta=12^\circ
  \rightarrow 16^\circ$. (d) Piecewise linear--hardening relationship
  of the effective shear stress $\tau~(\mathrm{N/mm^2})$ versus the
  normalized shearing displacement $\delta_\mathrm{II}/t$ for
  $\beta=16^\circ$. Properties of ICI APC-2/AS4 composite fibres and
  configuration and the range for $\beta$ were taken from
  \protect\cite{Kyriakides95}.}
  \label{fig:Kyr_beta}
\end{figure}

Figure \ref{fig:Kyr_beta}
shows numerical results from the current model using the properties
defined in Table \ref{tab:prop_Kyriakides} with $\beta$ values as
found in the published results. Note that the nondimensional total
end-shortening $\tilde{\Delta}$ is defined thus:
\begin{equation}
  \label{eq:Delta}
  \tilde{\Delta} = \tilde{\delta} + \tilde{b}(1-\cos\alpha).
\end{equation}
The actual kink band widths in the 5 tests were reported to range from
11 to 36 fibre diameters (directly corresponding to $\tilde{b}$ in the
current model) and the compressive strengths were found to average at
$1.119~\mathrm{kN/mm^2}$ with a standard deviation of
$0.043~\mathrm{kN/mm^2}$ (directly corresponding to $p$ in the current
model). The results from the current model show highly unstable
snap-back and hence the critical load would never be reached
realistically, see Figure \ref{fig:Kyr_beta}(a) and (b); a well
established feature for systems of this type
\cite{BudianskyFleck93}. For comparison purposes, the pressure $p$ is
taken at the point at which the structure stabilizes and reaches a
plateau; for the range of the $\beta$ angles considered, the
nondimensional stabilization pressure $\tilde{p}$ ranges from $80.5
\rightarrow 82.3$ which converts to an actual stabilization pressure
$p$ ranging from $0.969~\mathrm{kN/mm^2} \rightarrow
0.992~\mathrm{kN/mm^2}$: an error against the average from the
experimental results of between $11\% \rightarrow 13\%$, which is
sufficiently small to offer encouragement.

Of further interest is the comparison for the kink band width between
the tests and the current model. Observing the graph shown in Figure
\ref{fig:Kyr_beta}(c), as the kink band angle $\alpha$ increases,
initially the nondimensional kink band width $\tilde{b}$ falls from a
large value to a small value, approximately $5.6$ when
$\alpha=0.024~\mathrm{rad} (\approx 1.4^\circ)$. As $\alpha$ increases
further, the kink band width begins to increase slowly; see Table
\ref{tab:compare_Kyriakides} for details of some key points.
\begin{table}
  \centering
  \renewcommand\arraystretch{1.20}
  \begin{tabular}[htb]{c|c|c}
    Kink band angle & Case: $\beta=12^\circ$ & Case:
    $\beta=16^\circ$\\
    \hline
    $\alpha=\beta$ & $\tilde{b}=10.4$ & $\tilde{b}=10.2$\\
    $\alpha=2\beta$ & $\tilde{b}=17.0$ & $\tilde{b}=19.5$\\
  \end{tabular}
  \caption{Nondimensional kink band width values from the current
    model at different stages of deformation. The conditions
    $\alpha=\beta$ and $\alpha=2\beta$ are the points where the
    dilation within the band are effectively maximized and minimized
    respectively; experiments in \protect\cite{Kyriakides95} reported
    $\tilde{b}=11 \rightarrow 36$.}
  \label{tab:compare_Kyriakides}
\end{table}
According to the sequence described in Figures
\ref{fig:kinkgeol_seq}(b)--(d) the kink band itself maximizes dilation
when $\alpha=\beta$, minimizes it when $\alpha=2\beta$ and locks up
when $\alpha > 2\beta$. The results of the current model, particularly
when $\alpha=2\beta$, lie at the lower end of the range of observed
values of the band widths from the published experiments. This seems
sensible given that the lock-up condition used, where $\alpha=2\beta$,
represents a lower bound \cite{MAW_jmps04}, which implies that the
current model would also tend to predict lower bound kink band
widths. Hence, the results from the comparisons between the current
model and the published experiments \cite{Kyriakides95} are highly
encouraging; they offer very good quantitative agreement for the
loading and the geometric deformation -- key quantities that define
the kink band phenomenon.

\subsection{Parametric studies and discussion}

The favourable comparisons between the current model with the
published experiments in \cite{Kyriakides95} imply that the
fundamental physics of the system are captured by the current
approach. The study is therefore extended to present a series of model
parametric variations to establish their relative effects. The basic
geometric and material configuration is identical to that used in the
validation study presented in Table
\ref{tab:prop_Kyriakides}. Material and geometric parameters are
varied individually, while maintaining the remaining ones at their
original values. The parameters that are varied are the kink band
orientation angle $\beta$, the critical kink band angle,
$\alpha_\mathrm{C}$, the composite direct and shear moduli, $E_{11}$,
$E_{22}$ and $G_{12}$, and the shape of the piecewise linear
relationship for shear.

\subsubsection{Orientation angle}

\begin{figure}[htb]
\centering
\psfig{figure=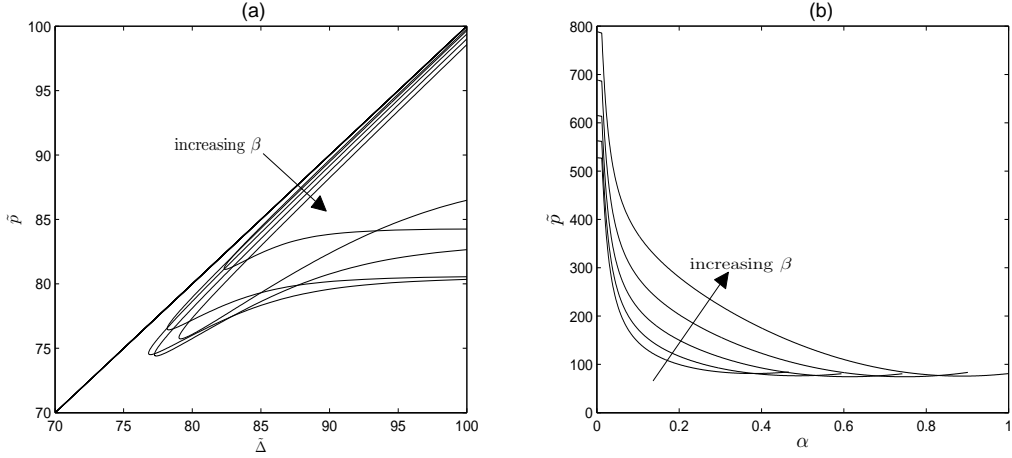,width=135mm}
\vspace{-3mm}
\caption{Equilibrium paths for different $\beta = 10^\circ \rightarrow
  30^\circ$ through nondimensional plots of load $\tilde{p}$ versus
  (a) total end-shortening $\tilde{\Delta}$ and (b) kink band angle
  $\alpha~(\mathrm{rad})$.}
  \label{fig:parambeta}
\end{figure}

\begin{figure}[htb]
\centering
\psfig{figure=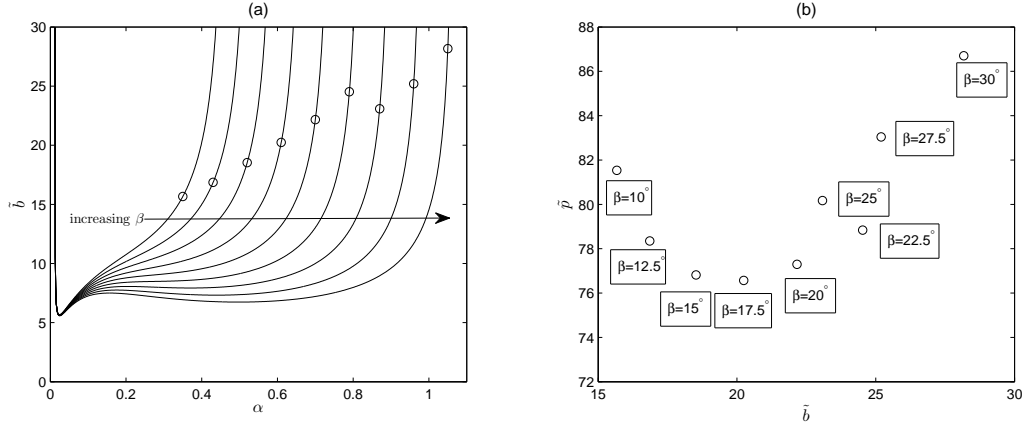,width=135mm}
\vspace{-3mm}
\caption{(a) Nondimensional kink band widths $\tilde{b}$ versus the
  kink band angle $\alpha~(\mathrm{rad})$ for a range of orientation angles
  $\beta=10^\circ \rightarrow 30^\circ$; circles mark the the lower
  bound lock-up condition $\alpha=2\beta$. (b) Values of
  nondimensional kink band widths $\tilde{b}$ and applied axial
  pressure $\tilde{p}$ at the lower bound lock-up condition.}
  \label{fig:betavar2}
\end{figure}

\begin{figure}[htb]
\centering
\psfig{figure=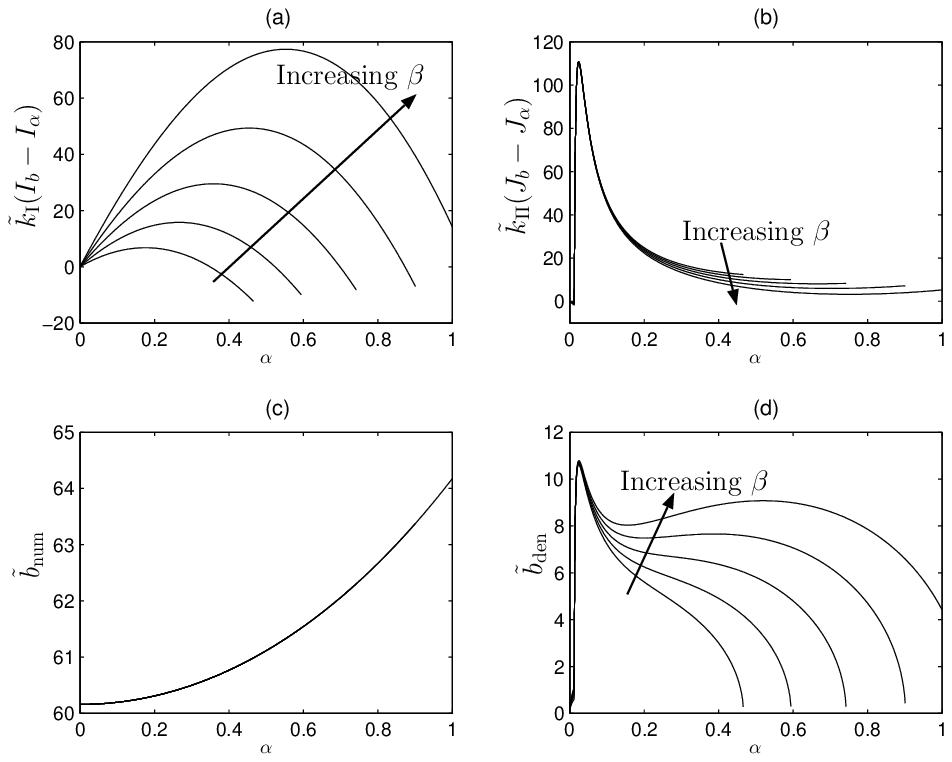,width=110mm}
\caption{Graphs of various terms from the expression for the
  nondimensional kink band width $\tilde{b}$, equation
  (\protect\ref{eq:bandwidth}) for $\beta=10^\circ \rightarrow
  30^\circ$ versus the kink band angle
  $\alpha~(\mathrm{rad})$. (a)--(b) Plots of dilation and shear terms
  respectively. (c)--(d) Plots of the numerator and denominator of the
  $\tilde{b}$ expression.}
  \label{fig:betavar_bnumden}
\end{figure}

In the current model, the orientation angle $\beta$ needs to be fixed
\emph{a priori}, hence the effects of different starting conditions
for the model need to be established. Increasing $\beta$ from
$10^{\circ}$ to $30^{\circ}$, a range that is representative of
laminate experiments in the literature
\cite{Kyriakides95,Gutkin2_CST_2010}, leads to a stiffer response for
increasing $\alpha$, as shown in Figure~\ref{fig:parambeta},
with the pressure capacity for $\beta=30^{\circ}$ being more than
double the capacity for $\beta=10^\circ$ for values of
$\alpha<\beta$. The graphs in Figure \ref{fig:betavar2}
raise an interesting point about the response particularly when $\beta
\geqslant 22.5^\circ$, which seems to define a boundary where the kink
band width $b$ loses its monotonically increasing property after it
initially troughs for a small value of $\alpha$, which was identified
as approximately $1.4^\circ$ in \S\ref{sec:validation}. Both graphs
show that the kink band width at the lower bound lock-up condition
$\alpha=2\beta$ temporarily peak when $\beta=22.5^\circ$. For higher
orientation angles the kink band width $b$ in fact peaks beyond
$\alpha=1.4^\circ$, then troughs and then resumes the monotonic rise
as seen for $\beta \leqslant 22.5^\circ$. Moreover, this also explains
the reason why the stabilization pressure increases for $\beta >
22.5^\circ$, as shown in Figure \ref{fig:parambeta}(a), since the
pressure has an inverse square relationship with the kink band width
as shown in the equilibrium equations (\ref{eq:equil_L_alpha}) and
(\ref{eq:equil_L_b}). The graphs presented in Figure
\ref{fig:betavar_bnumden}
attribute this loss of monotonicity in $b$ (beyond $\alpha=1.4^\circ$)
to the dominating influence of the dilation terms for larger $\beta$,
particularly in the region of maximum dilation where $\alpha \approx
\beta$. In the first instance, it should be recalled that when $\beta$
is larger the potential maximum dilation displacement
$\delta_\mathrm{I}$ is also larger relative to $\delta_\mathrm{II}$
when $\alpha=\beta$. Figure \ref{fig:betavar_bnumden}(a) shows that
the maximum of the dilation term $\tilde{k}_\mathrm{I} (I_b-I_\alpha)$
from the expression for $b$, \emph{i.e.}\ equation
(\ref{eq:bandwidth}), increases substantially with $\beta$ whereas
Figure \ref{fig:betavar_bnumden}(b) shows only very marginal changes
in the respective shear term $\tilde{k}_\mathrm{II}
(J_b-J_\alpha)$. The numerator in equation (\ref{eq:bandwidth}),
$\tilde{b}_\mathrm{num}$, which represents the influence of bending,
is independent of $\beta$, as shown in Figure
\ref{fig:betavar_bnumden}(c) but the respective denominator,
$\tilde{b}_\mathrm{den}$, shows that the dilation term influences the
values significantly for the higher $\beta$ values, as shown in Figure
\ref{fig:betavar_bnumden}(d). Once $\alpha$ gradually increases above
$\beta$ the dilation displacement progressively reduces and the shear
term begins to dominate with the result that the kink band width
resumes growth and lock-up occurs. This effect is similar to that
found in the geological model with the introduction of the foundation
spring of stiffness $k_f$ \cite{MAW_jmps04}, as shown in Figure
\ref{fig:kinkgeol_2layers}; the kink band width was also found to
plateau with higher foundation stiffnesses. It is worth noting that if
destiffening in the constitutive law for dilation was introduced that the
effect found in the present case would be generally less pronounced.

\subsubsection{Critical shear angle and modulus}

\begin{figure}[htb]
\centering
\psfig{figure=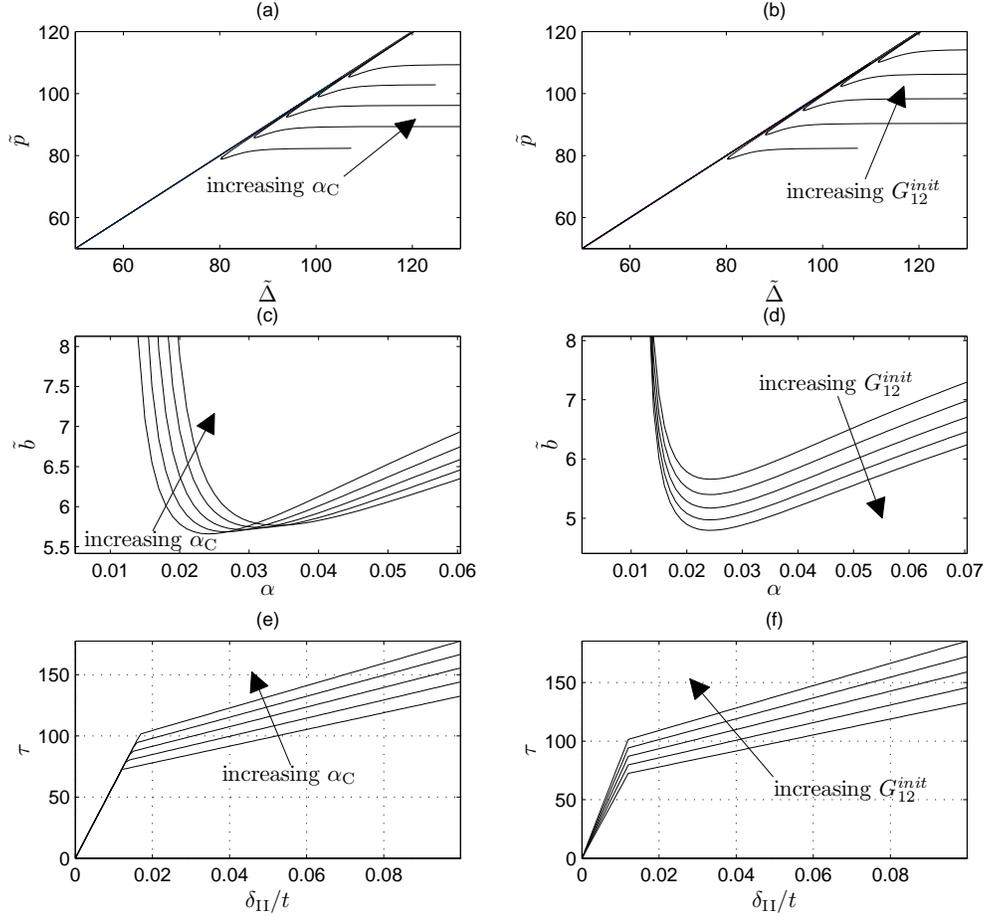,width=130mm}
\caption{Comparison of the response of the kink band formation for
  different $\alpha_\mathrm{C}$ values in (a), (c) and (e) for the
  range of $\alpha_\mathrm{C}= 0.69^{\circ}\rightarrow 0.96^{\circ}$
  and initial $G_{12}$ values from $6.03~\mathrm{kN/mm^2} \rightarrow
  8.45~\mathrm{kN/mm^2}$ (in (b), (d) and (f). Nondimensional plots of
  (a)--(b) load $\tilde{p}$ versus total end-shortening
  $\tilde{\Delta}$, (c)--(d) kink band width $\tilde{b}$ versus kink
  band angle $\alpha~(\mathrm{rad})$.  (e)--(f) show the relationship
  of the effective shear stress $\tau$ versus the normalized shearing
  displacement $\delta_\mathrm{II}/t$. Note that $\beta=12^\circ$
  throughout. }
  \label{fig:paramacg12}
\end{figure}

Increasing the critical kink band angle from $\alpha_\mathrm{C} =
0.69^{\circ} \rightarrow 0.96^{\circ}$ (with a fixed limiting
displacement $\delta_\mathrm{M}/t=-0.094$ as before) shows an
increase in the critical shear stress before destiffening occurs -- see
Figure~\ref{fig:paramacg12}(e) -- and leads to a monotonic increase of
the axial pressure $p$ and the minimum kink band width $b$ -- see
Figures~\ref{fig:paramacg12}(a) and (c).
A subtly different pattern is observed in
Figures~\ref{fig:paramacg12}(b) and (d) where trends for increasing
the initial shear modulus ($G_{12}^\mathrm{init}=6.03
\mathrm{kN}/\mathrm{mm^2} \rightarrow 8.45\mathrm{kN}/\mathrm{mm^2}$),
lead to higher stabilization pressures but smaller minimum kink band
widths.  These are logical results since the effect of increasing the
critical kink band angle will lead to a later destabilization in shear
and hence increase the load and band width; the increase in the
initial shear modulus increases the resistance against shearing -- the
process of kink banding therefore requires more axial pressure to
overcome this increased stiffness. However, the increased shear
stiffness reduces the kink band width since there is a greater
resistance to that type of deformation.

\subsubsection{Hardening and softening in shear}

\begin{figure}[htb]
\centering
\psfig{figure=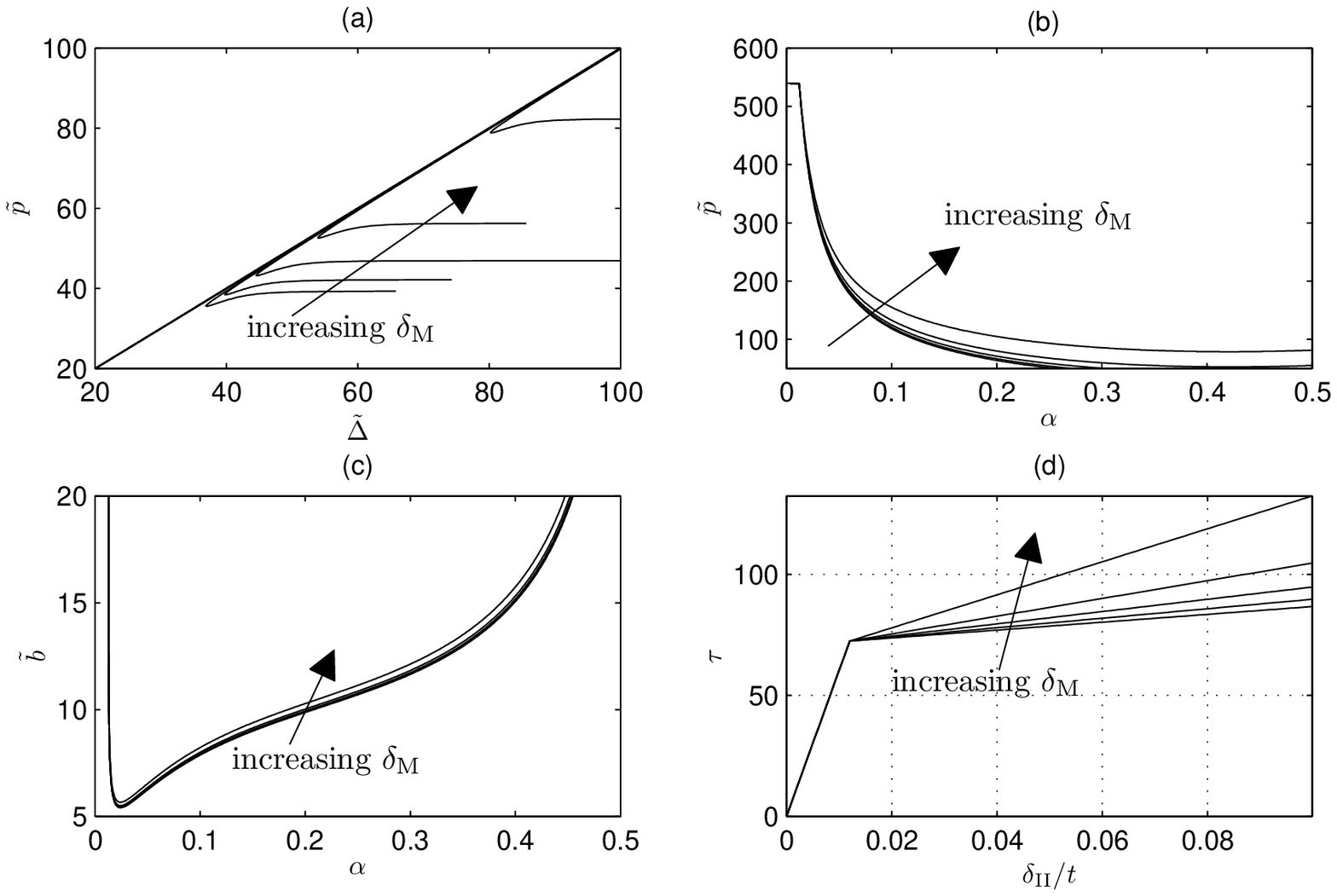,width=115mm}
\caption{Nondimensional plots of load $\tilde{p}$ versus (a) total
  end-shortening $\tilde{\Delta}$ and (b) kink band angle
  $\alpha~(\mathrm{rad})$; (c) kink band width $\tilde{b}$ versus kink
  band angle $\alpha~(\mathrm{rad})$; (d) Piecewise linear--hardening
  relationship of the effective shear stress $\tau~(\mathrm{N/mm^2})$
  versus the normalized shearing displacement $\delta_\mathrm{II}/t$.
  Range of $\delta_\mathrm{M}/t=-0.435 \rightarrow -0.094$ and
  $\beta=12^\circ$.}
  \label{fig:pwhard}
\end{figure}

\begin{figure}[htb]
\centering
\psfig{figure=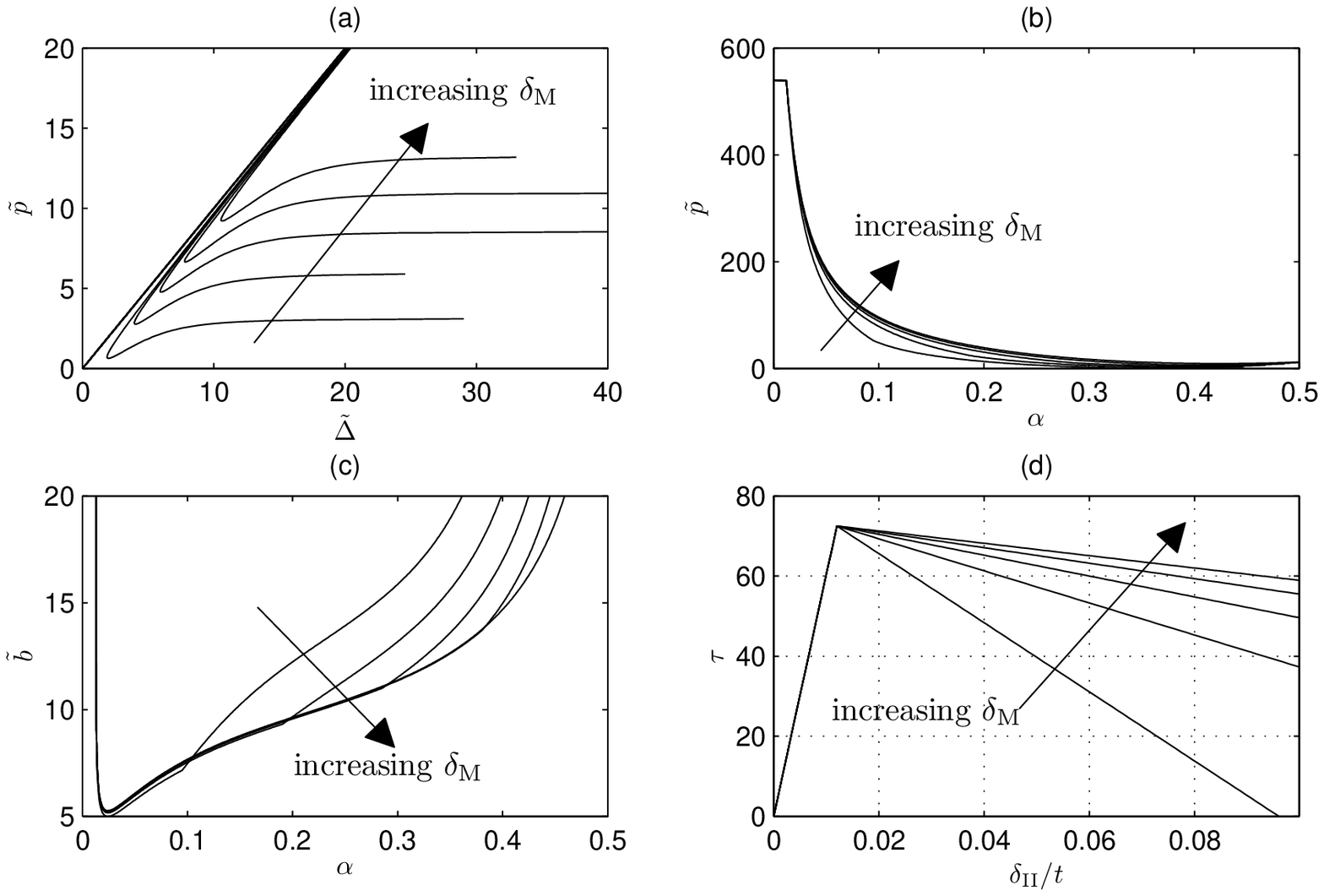,width=115mm}
\caption{Nondimensional plots of load $\tilde{p}$ versus (a) total
  end-shortening $\tilde{\Delta}$ and (b) kink band angle
  $\alpha~(\mathrm{rad})$; (c) kink band width $\tilde{b}$ versus kink
  band angle $\alpha~(\mathrm{rad})$; (d) Piecewise linear--softening
  relationship of the effective shear stress $\tau~(\mathrm{N/mm^2})$
  versus the normalized shearing displacement $\delta_\mathrm{II}/t$.
  Range of $\delta_\mathrm{M}/t= 0.096 \rightarrow 0.482$ and
  $\beta=12^\circ$.}
  \label{fig:pwsoft}
\end{figure}

The variation in the piecewise linear model for the shearing response
is now discussed. The constitutive behaviour, $F_\mathrm{II}$ versus
$\delta_\mathrm{II}$ has been hitherto assumed to be a
linear--hardening law which corresponded with the data from the
literature used in the validation exercise. Figure \ref{fig:pwhard}
shows results for different secondary slopes while they remain
positive (a linear--hardening law). Figure \ref{fig:pwsoft}
shows results for reducing the secondary slope further such that they
become negative (a linear--softening law). The results exhibit
progressive behaviour; the reduced secondary slopes reduce the load
carrying capacity but increase the kink band widths. This can be
understood from the softening of the internal structure giving less
resistance to the kink banding process, allowing for larger rotations
and gross deformations. For the cases presented in Figure
\ref{fig:pwsoft}, the negative secondary slope mimics the behaviour of
a fracture process where the shear stiffness and strength has vanished
and mode II fracture and crack propagation would occur. However, as
described above, a similar pattern remains with the strength reducing
and the band widths increasing for weaker properties in shear, which
appears to be entirely logical. The detailed effects of crack
propagation have been left for future work although recent work on
buckling-driven delamination \cite{imamat_christina} has suggested
that an analytical treatment of such effects may indeed be tractable.

\subsubsection{Young's moduli}

\begin{figure}[htb]
\centering
\psfig{figure=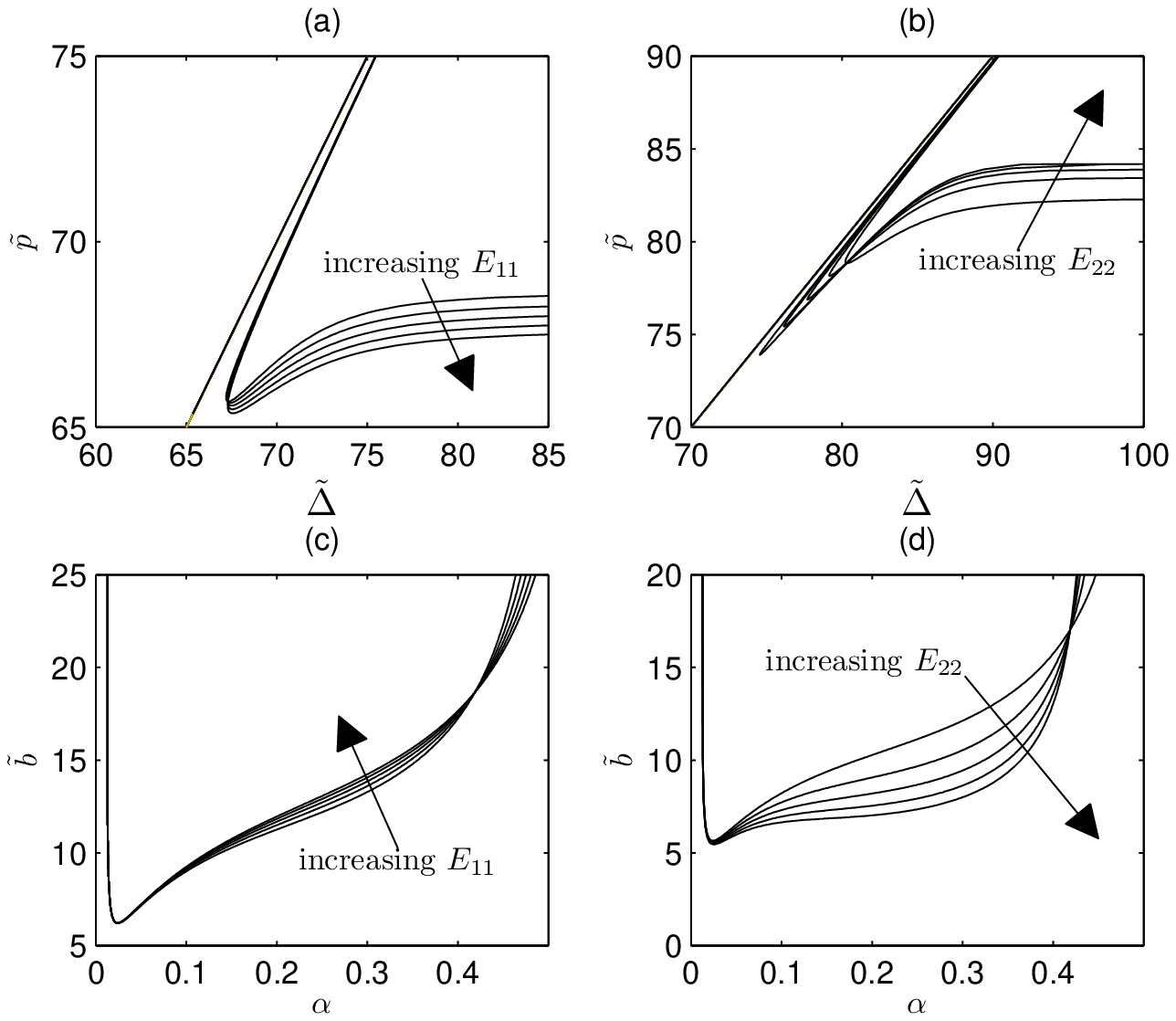,width=105mm}
\caption{Nondimensional plots for the range of axial direct modulus
  $E_\mathrm{11}= 130.7~\mathrm{kN}/\mathrm{mm^2} \rightarrow
  261.5~\mathrm{kN}/\mathrm{mm^2}$ in (a) and (c) and transverse direct
  modulus $E_\mathrm{22}= 10.4~\mathrm{kN}/\mathrm{mm^2} \rightarrow
  52~\mathrm{kN}/\mathrm{mm^2}$ in (b) and (d). (a)--(b) Load
  $\tilde{p}$ versus $\alpha$; (c)--(d) kink band width $\tilde{b}$
  versus the kink band angle $\alpha~(\mathrm{rad})$. Note that
  $\beta=12^\circ$ throughout.}
  \label{fig:E22}
\end{figure}

Results for a two-fold increase in the axial modulus $E_{11}$ suggest
that this only has a marginal effect on the stabilization pressure
($\approx 1.5\%$ increase), whereas increasing the lateral modulus
$E_{22}$ results in a significantly stiffer response; the system
stabilizing to a smaller kink band width (see Figure~\ref{fig:E22}).
These, again, are logical results since the effect of increasing the
lateral modulus $E_{22}$ increases the resistance against dilation;
the process of kink banding therefore requiring more axial pressure to
overcome this. Increasing the axial modulus increases the axial
stiffness $k$, which in turn effectively reduces the relative dilation
and shear stiffnesses without affecting the relative bending
stiffness, see the scaling relationships in equation
(\ref{eq:nondim}). Since bending is currently assumed to be purely
linear, its relative effect becomes progressively more pronounced and
then outweighs the reduced dilation and shear effects at large
rotations. Obviously, if the bending was assumed to plateau due to
plasticity, this effect would be limited.

\section{Concluding remarks}

An analytical, nonlinear, potential energy based model for kink
banding in compressed unidirectional laminated composite panels has
been presented. Comparisons of results with published experiments from
the literature suggest that very good agreement can be achieved from
this relatively simple mechanical approach provided certain important
characteristics are incorporated:
\begin{enumerate}
\item \emph{Interlamina dilation and shearing}: the kink band rotation
  naturally causes shearing and changes the gap between the laminae,
  the matrix within the composite needs to resist both of these
  displacements for the laminate to have integrity and significant
  structural strength.
\item \emph{Bending energy}: the resistance to rotation sets a length
  scale which in this case is the kink band width $b$.
\end{enumerate}
Linear constitutive relationships for the mechanisms of bending and
dilation, and piecewise linear for the process of shearing, together
with nonlinear geometric relationships have been applied. The approach
has been successful such that the mechanical response captures the
fundamental physics of kink banding and agrees with the experiments
from \cite{Kyriakides95} in terms of kink band widths and loading
levels without having to resort to sophisticated numerical or
continuum formulations. Unlike the geological model \cite{MAW_jmps04},
where a relationship was derived for the band orientation $\beta$ that
was related to the overburden pressure $q$, in the current case the
angle $\beta$ has to be assumed \emph{a priori} since, as far as the
authors are aware, no satisfactory procedure for predicting $\beta$
for composite laminates exists. For laminates, the magnitude of the
orientation angle $\beta$ has been largely attributed to the
manufacturing process \cite{Kyriakides95,Budiansky83}.  However, if
the overburden pressure is considered to be the controlling parameter
for the equivalent ``manufacturing process'' that keeps the geological
layers behaving together, then future work on modelling the process of
manufacturing composite laminates may bear fruit; an indication of the
parameters that govern the orientation angle for the current case may
be established. Although this is a shortcoming for the present model,
the results from the parametric study are very encouraging with the
trends appearing to be entirely logical.

The current model can of course be used as a basis for further work.
In the first instance, the formation of new kink bands can be
investigated since work derived from similar approaches to the current
one exist for the geological model \cite{MAW_jmps05,edmunds_jsg10}. In
particular, in \cite{MAW_jmps05} the lock-up criterion was employed as
the condition to introduce a new kink band; although it was assumed
that the original kink band stops growing, the comparisons between
theory and experiments were shown to be very good. The piecewise
linear formulation applied currently for shear could also be extended
to include dilation giving the possibility of mixed mode fracture
\cite{Hutch92} for the first kink band. Moreover, loading cases that
are more complex than uniform compression could be investigated; for
example, numerical approaches have been developed in
\cite{KyriakidesFE,Gutkin1_CST_2010} with varying degrees of success
to investigate the formation of kink bands where there is a
combination of shear and compression.  An additional complication in
the combined loading case is that the kink band propagation across the
sample tends to occur more gradually in contrast to the present case
where the formation process is fast.

\bibliography{allrefs}

\begin{thebibliography}{10}

\bibitem{Anderson}
Anderson, T.~B.
\newblock Kink-bands and related geological structures.
\newblock {\em Nature}, 202:272--274, 1964.

\bibitem{HoMeWi}
Hobbs, B.~E., Means, W.~D., and Williams, P.~F.
\newblock {\em An Outline of Structural Geology}.
\newblock Wiley, New York, 1976.

\bibitem{PrCo}
Price, N.~J. and Cosgrove, J.~W.
\newblock {\em Analysis of geological structures}.
\newblock Cambridge University Press, Cambridge, 1990.

\bibitem{BudianskyFleck93}
Budiansky, B. and Fleck, N.~A.
\newblock Compressive failure of fiber composites.
\newblock {\em {J.\ Mech.\ Phys.\ Solids}}, 41:183--211, 1993.

\bibitem{Kyriakides95}
Kyriakides, S., Arseculeratine, R., Perry, E.~J., and Liechti, K.~M.
\newblock On the compressive failure of fiber reinforced composites.
\newblock {\em {Int.\ J.\ Solids Struct.}}, 32:689--738, 1995.

\bibitem{ReidPeng97}
Reid, S.~R. and Peng, C.
\newblock Dynamic uniaxial crushing of wood.
\newblock {\em {Int. J. Impact Eng.}}, 19:531--570, 1997.

\bibitem{KyriakidesExpt}
Vogler, T.~J. and Kyriakides, S.
\newblock On the initiation and growth of kink bands in fiber composites. {Part
  I:} experiments.
\newblock {\em {Int.\ J.\ Solids Struct.}}, 38:2639--2651, 2001.

\bibitem{Byskov2002}
Byskov, E., Christoffersen, J., Christensen, C.~D., and Poulsen, J.~S.
\newblock Kinkband formation in wood and fiber composites---morphology and
  analysis.
\newblock {\em {Int.\ J.\ Solids Struct.}}, 39:3649--3673, 2002.

\bibitem{DaSilva20078685}
Da~Silva, A. and Kyriakides, S.
\newblock Compressive response and failure of balsa wood.
\newblock {\em {Int.\ J.\ Solids Struct.}}, 44(25--26):8685--8717, 2007.

\bibitem{Pimenta1_CST_2009}
Pimenta, S., Gutkin, R., Pinho, S.~T., and Robinson, P.
\newblock A micromechanical model for kink-band formation: Part
  {I}---experimental study and numerical modelling.
\newblock {\em {Compos. Sci. Technol.}}, 69:948--955, 2009.

\bibitem{Hobbs_textile}
Hobbs, R.~E., Overington, M.~S., Hearle, J.~W.~S., and Banfield, S.~J.
\newblock Buckling of fibres and yarns within ropes and other fibre assemblies.
\newblock {\em {J. Textile Inst.}}, 91(3):335--358, 2000.

\bibitem{MAW_cst}
Edmunds, R. and Wadee, M.~A.
\newblock On kink banding in individual {PPTA} fibres.
\newblock {\em {Compos. Sci. Technol.}}, 65(7--8):1284--1298, 2005.

\bibitem{Rosen}
Rosen, B.~W.
\newblock Mechanics of composite strengthening.
\newblock In Bush, S.~H., editor, {\em Fiber Composite Materials}, pages
  37--75. American Society of Metals, 1965.

\bibitem{Argon}
Argon, A.~S.
\newblock Fracture of composites.
\newblock {\em {Treatise Mater. Sci. Technol.}}, 1:79--114, 1972.

\bibitem{Budiansky83}
Budiansky, B.
\newblock Micromechanics.
\newblock {\em {Comput.\ \& Struct.}}, 16:3--12, 1983.

\bibitem{FuZhang06}
Fu, Y.~B. and Zhang, Y.~T.
\newblock Continuum-mechanical modelling of kink-band formation in fibre
  reinforced composites.
\newblock {\em {Int.\ J.\ Solids Struct.}}, 43(11--12):3306--3323, 2006.

\bibitem{KyriakidesFE}
Vogler, T.~J., Hsu, S.-Y., and Kyriakides, S.
\newblock On the initiation and growth of kink bands in fiber composites. {Part
  II:} analysis.
\newblock {\em {Int.\ J.\ Solids Struct.}}, 38:2653--2682, 2001.

\bibitem{BudianskyFleckAmazigo98}
Budiansky, B., Fleck, N.~A., and Amazigo, J.~C.
\newblock On kink-band propagation in fiber composites.
\newblock {\em {J.\ Mech.\ Phys.\ Solids}}, 46:1637--1653, 1998.

\bibitem{Pimenta2_CST_2009}
Pimenta, S., Gutkin, R., Pinho, S.~T., and Robinson, P.
\newblock A micromechanical model for kink-band formation: Part
  {II}---analytical modelling.
\newblock {\em {Compos. Sci. Technol.}}, 69:956--964, 2009.

\bibitem{Fleck97}
Fleck, N.~A.
\newblock Compressive failure of fiber composites.
\newblock {\em {Adv. Appl. Mech.}}, 33:43--117, 1997.

\bibitem{MAW_jsg}
Hunt, G.~W., Peletier, M.~A., and Wadee, M.~A.
\newblock The {Maxwell} stability criterion in pseudo-energy models of kink
  banding.
\newblock {\em {J.\ Struct.\ Geol.}}, 22(5):669--681, 2000.

\bibitem{MAW_jmps04}
Wadee, M.~A., Hunt, G.~W., and Peletier, M.~A.
\newblock Kink band instability in layered structures.
\newblock {\em {J.\ Mech.\ Phys.\ Solids}}, 52(5):1071--1091, 2004.

\bibitem{MAW_jmps05}
Wadee, M.~A. and Edmunds, R.
\newblock Kink band propagation in layered structures.
\newblock {\em {J.\ Mech.\ Phys.\ Solids}}, 53(9):2017--2035, 2005.

\bibitem{Gutkin2_CST_2010}
Gutkin, R., Pinho, S.~T., Robinson, P., and Curtis, P.~T.
\newblock On the transition from shear-driven fibre compressive failure to
  fibre kinking in notched {CFRP} laminates under longitudinal compression.
\newblock {\em {Compos. Sci. Technol.}}, 70:1223--1231, 2010.

\bibitem{imamat_christina}
Wadee, M.~A. and V\"ollmecke, C.
\newblock Semi-analytical modelling of buckling driven delamination in
  uniaxially compressed damaged plates.
\newblock {\em {IMA J.\ Appl.\ Math.}}, 76(1):120--145, 2011.

\bibitem{edmunds_jsg10}
Edmunds, R., Hicks, B.~J., and Mullineux, G.
\newblock Drawing parallels: Modelling geological phenomena using constraint
  satisfaction.
\newblock {\em {J.\ Struct.\ Geol.}}, 32(7):997--1008, 2010.

\bibitem{Hutch92}
Hutchinson, J.~W. and Suo, Z.
\newblock Mixed mode cracking in layered materials.
\newblock {\em {Adv. Appl. Mech.}}, 29:63--191, 1992.

\bibitem{Gutkin1_CST_2010}
Gutkin, R., Pinho, S.~T., Robinson, P., and Curtis, P.~T.
\newblock Micro-mechanical modelling of shear-driven fibre compressive failure
  and of fibre kinking for failure envelope generation in {CFRP} laminates.
\newblock {\em {Compos. Sci. Technol.}}, 70:1214--1222, 2010.

\end{thebibliography}
\end{document}